\documentclass[article,superscriptaddress,twocolumn]{revtex4}
\usepackage{amssymb,amsfonts,amsmath}
\usepackage{graphicx}
\usepackage{booktabs}
\usepackage{units}

\begin{document}

\title{Robust circadian clocks from coupled protein modification and
  transcription-translation cycles}

\author{David Zwicker}
\affiliation{FOM Institute for Atomic and Molecular
    Physics (AMOLF), Science Park 104, 1098 XG Amsterdam, The
    Netherlands}
\author{David K. Lubensky}
\affiliation{Department of Physics, University of Michigan, Ann Arbor,
  MI 48109-1040}
\author{Pieter Rein ten Wolde}
\affiliation{FOM Institute for Atomic and Molecular
    Physics (AMOLF), Science Park 104, 1098 XG Amsterdam, The
    Netherlands}

\begin{abstract}
  The cyanobacterium \textit{Synechococcus elongatus} uses both a
  protein phosphorylation cycle and a transcription-translation cycle
  to generate circadian rhythms that are highly robust against
  biochemical noise. We use stochastic simulations to analyze how
  these cycles interact to generate stable rhythms in growing,
  dividing cells. We find that a protein phosphorylation cycle by itself is
  robust when protein turnover is low.  For high decay or dilution rates (and co
mpensating synthesis rate), however, the phosphorylation-based oscillator
  loses its integrity.  Circadian rhythms thus cannot be generated
  with a phosphorylation cycle alone when the growth rate, and
  consequently the rate of protein dilution, is high enough; in
  practice, a purely post-translational clock ceases to function well
  when the cell doubling time drops below the 24 hour clock period.
  At higher growth rates, a transcription-translation cycle becomes
  essential for generating robust circadian rhythms. Interestingly,
  while a transcription-translation cycle is necessary to sustain a
  phosphorylation cycle at high growth rates, a phosphorylation cycle
  can dramatically enhance the robustness of a
  transcription-translation cycle at lower protein decay
  or dilution rates. Our analysis thus predicts that both cycles are
  required to generate robust circadian rhythms over the full range
  of growth conditions. 
\end{abstract}

\maketitle

\section{Introduction}
Many organisms use circadian clocks to anticipate changes between day
and night \cite{Johnson:2008jy}. It had long been believed that these
clocks are driven primarily by transcription-translation cycles (TTCs)
built on negative feedback. However, while some circadian clocks can
maintain robust rhythms for years in the absence of any daily cue
\cite{Johnson:2008jy}, recent experiments have vividly demonstrated
that gene expression is often highly stochastic \cite{Elowitz02}. This
raises the question of how these clocks can be robust against
biochemical noise.  In multicellular organisms, the robustness might
be explained by intercellular interactions
\cite{Liu:1997kb,Yamaguchi:2003vl}, but it is now known that even
unicellular organisms can have very stable circadian rhythms. The
clock of the cyanbacterium {\em Synechococcus elongatus}, for example,
has a correlation time of several months \cite{Mihalcescu:2004yq},
even though the clocks of the different cells in a population hardly
interact with one another
\cite{Mihalcescu:2004yq,Amdaoud:2007rc}. Interestingly, it has
recently been discovered that the {\em S. elongatus} clock also
includes a protein phosphorylation cycle (PPC) that can run
independently of the TTC \cite{Tomita:2005bn,Nakajima:2005gq}. It has
been suggested that this protein modification oscillator
\cite{Nakajima:2005gq}, possibly in combination with the
transcription-translation oscillator \cite{Kitayama:2008kz}, is
responsible for the clock's striking noise resistance.  Here, we use
mathematical modeling to study how these two clocks interact in
growing, dividing cells. We find that the PPC alone is very stable
when cell growth is negligible, but it begins to fail as the growth
rate increases, and for doubling times typical of
\textit{S. elongatus} it cannot function without help from a TTC.  We
then use stochastic simulations to show that a PPC can, however,
dramatically enhance the robustness of a TTC, especially when cells
are growing slowly.  The two mechanisms thus perform best in
complementary situations, and it is likely that both oscillators are
necessary to maintain even minimal clock stability across the full
range of conditions and growth rates that bacteria encounter in the
wild.  Importantly, the TTC and the PPC in \textit{S. elongatus} are
much more tightly intertwined than conventional coupled phase
oscillators; as a result, 
the combination of the two far outperforms not just each of its two
components individually, but also a hypothetical system in which the
two parts are coupled in normal textbook fashion~\cite{Syncbook}.

In {\em S. elongatus}, the central components of the clock are the
three genes {\em kaiA}, {\em kaiB}, and {\em kaiC}
\cite{Ishiura:1998ay}.  Under continuous light conditions, the levels of mRNA from the {\em kaiBC} operon
and of the protein KaiC oscillate in a circadian fashion with a 6 hour
lag between their peaks
\cite{Xu:2000zg}.  Overexpression of phosphorylated KaiC abolishes
{\em kaiBC} expression \cite{Nakahira:2004qz,Nishiwaki:2004kc}, while
a transient increase in KaiC resets the phase of the oscillator
\cite{Xu:2000zg}. These observations led to the proposal that the Kai
system is a transcription-translation oscillator, with KaiC negatively
regulating its own transcription. In 2005, however, Kondo and
coworkers showed that KaiC is phosphorylated in a cyclical manner
with a period of 24 hours, even when {\em kaiBC} transcription is
inhibited \cite{Tomita:2005bn}. Still more remarkably, the rhythmic
phosphorylation of KaiC could be reconstituted in the test tube in the
presence of only KaiA, KaiB, and ATP \cite{Nakajima:2005gq}. This
raised the possibility that the principal pacemaker of the clock is
not a transcription-translation cycle, but a protein phosphorylation
cycle \cite{Nakajima:2005gq}. Yet, in 2008, the same group showed that
circadian oscillations of gene expression persist even when KaiC is
always held in a highly phosphorylated state
\cite{Kitayama:2008kz}. They thus concluded that the clock is driven by both a TTC and a PPC, and suggested that the interactions between
the two oscillators may enhance the robustness of the clock
\cite{Kitayama:2008kz}.

The PPC has been
characterized experimentally in considerable detail. It is known, for example, that KaiC forms a
homo-hexamer \cite{Kageyama:2003ka}, KaiA a dimer
\cite{Kageyama:2003ka}, and KaiB a dimer
\cite{Kageyama:2003ka,Kageyama:2006jp} or a tetramer
\cite{Hitomi:2005ej}.  KaiC has two phosphorylation sites per protein
monomer, which are phosphorylated and dephosphorylated in a definite
sequence as a result of KaiC's autokinase and autophosphatase activity
\cite{Nishiwaki:2007ao,Rust:2007rb}. KaiA stimulates KaiC
phosphorylation \cite{Iwasaki:2002um,Xu:2003my}, while KaiB negates
the effect of KaiA
\cite{Iwasaki:2002um,Xu:2003my,Kitayama:2003ts,Williams:2002oh}.
Thanks to the wealth of available experimental data, the PPC has
proven an unusually fruitful system for mathematical modeling
\cite{Rust:2007rb,Takigawa-Imamura:2006ss,Emberly:2006kr,Mehra:2006zk,Zon:2007ly,Clodong:2007ff,Mori:2007pd,Miyoshi:2007uq,Yoda:2007iq,Eguchi:2008sw}.

\begin{figure}[t]
\center
\includegraphics[width=4.5cm,angle=0]{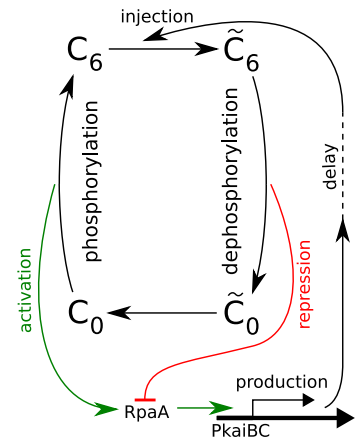}
\caption{A cartoon of how the TTC might interact with the PPC
  \cite{Taniguchi:2010tt}. Active RpA activates {\em kaiBC}
  expression. During the phosphorylation phase, KaiC activates RpaA,
  while during the dephosphorylation phase, KaiC represses
  RpaA. \label{fig:cartoon}}
\end{figure}

The TTC appears to encompass several different regulatory pathways and
is much less well understood.  Multiple lines of evidence indicate
that circadian promoter activity can be achieved without any specific
\textit{cis}-regulatory
element~\cite{Ito:2009dm,Vijayan:2009rq,Ditty:2003vy}, suggesting that
the clock may regulate transcription at least in part through changes
in chromosome compaction and
superhelicity~\cite{Vijayan:2009rq,Smith:2006hf,Woelfle:2007ws}.
Several proteins important for transcriptional regulation of the
\textit{kaiBC} operon have also been identified.  In particular, LabA
and the sensor histidine kinase CikA have been implicated in negative
feedback of phosphorylated KaiC on its own transcription during the
subjective
night~\cite{Ishiura:1998ay,Nakahira:2004qz,Nishiwaki:2004kc,Taniguchi:2007pf,Taniguchi:2010tt},
while the SasA histidine kinase, acting through the RpaA response
regulator, seems to mediate positive feedback during the subjective
day~\cite{Iwasaki:2000fl,Takai:2006cz}.  Epistasis analysis suggests
that LabA may likewise act through RpaA~\cite{Taniguchi:2007pf}.  The
picture that emerges is thus that during the phosphorylation phase of
the PPC, KaiC activates {\em kaiBC} expression, while in the
dephosphorylation phase, it represses {\em kaiBC} expression
\cite{Taniguchi:2010tt} (Fig. \ref{fig:cartoon}).

In this study, we perform stochastic simulations to investigate the
roles of the PPC and the TTC.  We first examine a situation in which
the total number of each Kai protein is {\em constant}---they are
neither produced nor destroyed---and only the PPC is operative.  We
show that in this case the PPC is highly robust against noise arising from the
intrinsic stochasticity of chemical reactions. Even for reaction
volumes smaller than the typical volume of a cyanobacterium, the
correlation time is longer than that observed experimentally
\cite{Mihalcescu:2004yq}.  Living cells, however, constantly grow and
divide, and proteins must thus be synthesized to balance dilution.  In
fact, dilution can be thought of as introducing an effective protein
degradation rate set by the cell doubling time. We therefore next
study a PPC in which the Kai proteins are produced and degraded with
rates that are constant in time. The simulations reveal that protein
synthesis and decay dramatically reduce the viability of the PPC; we predict that for a cell doubling time of 24 hours and a
bacterial volume of 1 $\mu\text{m}^{3}$, the PPC dephases in roughly
10 days, much faster than real \textit{S. elongatus}
\cite{Mihalcescu:2004yq}.  The same loss of oscillations is seen even
in the limit of infinite volume, indicating it is not due solely to
noise.  Instead, the constant synthesis of proteins, which we assume
are all initially created in the same phosphorylation state,
necessarily injects KaiC hexamers with the ``wrong'' phosphorylation
levels at certain phases of the cycle~\cite{Johnson:2008jy,Ito:2007ca}; if these
appear fast enough, they can destroy the oscillation.  One role of the
TTC is thus to introduce proteins only when the phosphorylation state
of the freshly made KaiC matches that of the PPC.  Our simulations of
a combined PPC and TTC reveal that a TTC can indeed greatly enhance the
robustness of the PPC, yielding correlation times
consistent with those measured experimentally
\cite{Mihalcescu:2004yq}. Finally, we consider whether the PPC is
needed at all, or whether one could build an equally good circadian
clock using only a TTC.  We find that it is possible to construct a
TTC with a period of 24 hours and the observed correlation time of a few months \cite{Mihalcescu:2004yq}. However, this
comes at the expense of very high protein synthesis and decay rates,
which impose an extra energetic burden on the cell. Our results thus
suggest that a PPC allows for a more robust oscillator at a lower
cost.  Although our models 
are simplified, 
we argue in the Discussion section that our qualitative results are
unavoidable consequences of the interaction between a circadian clock
and cell growth and so should hold far more generally.

\section{Results}
\subsection{A protein phosphorylation cycle with constant protein
  concentrations is highly robust}
In recent years, the PPC has been mathematically modelled with
considerable success
\cite{Rust:2007rb,Takigawa-Imamura:2006ss,Emberly:2006kr,Mehra:2006zk,Zon:2007ly,Clodong:2007ff,Mori:2007pd,Miyoshi:2007uq,Yoda:2007iq,Eguchi:2008sw}
(for a review, see \cite{Markson:2009rz}). All models endow KaiC with
an intrinsic ability to cyclically phosphorylate and dephosphorylate
itself, but they differ in how the cycles of the individual KaiC
proteins are synchronized. In this manuscript, we adopt the model
developed by us \cite{Zon:2007ly}, which is one of several
based on a mechanism that we called ``differential
affinity'' \cite{Rust:2007rb,Takigawa-Imamura:2006ss,Zon:2007ly,Clodong:2007ff}: KaiA stimulates KaiC phosphorylation,
but the limited supply of KaiA dimers binds preferentially to those
KaiC molecules that are falling behind in the cycle, allowing them to
catch up. Specifically, in our model each KaiC hexamer can switch
between an active conformational state ${{\rm C}}_i$, where the number
$i$ of phosphorylated monomers tends to increase, and an inactive state
$\widetilde{{\rm C}}_i$, where $i$ tends to decrease
(Fig. \ref{fig:cartoon}); KaiA stimulates phosphorylation of active
KaiC, but is sequestered by complexes containing KaiB and inactive
KaiC.  KaiC in the inactive state can thus delay 
the progress of fully
dephosphorylated hexamers that have already switched back to the
active state and are ready to be phosphorylated again.  With A and B
denoting, respectively, a KaiA dimer and a KaiB dimer, the model
becomes:
\begin{alignat}{1}
  &{\rm C}_i \underset{b_i}{\overset{f_i}{\rightleftarrows}}
  \widetilde{{\rm C}}_i,\,\,{\rm C}_i + {\rm A} \underset{k_i^{\rm Ab}}{\overset{k_i^{\rm
        Af}}{\rightleftarrows}} {\rm AC}_i \overset{k_{\rm
      pf}}{\rightarrow} {\rm C}_{i+1} + {\rm A} \label{eq:P0}\\
&\widetilde{\rm C}_i + {\rm B} \underset{\tilde k^{\rm Bb}_i}{\overset{2\tilde k^{\rm Bf}_i}{
\rightleftarrows}}
{\rm B}\widetilde{\rm C}_i, 
{\rm B}\widetilde{\rm C}_i + {\rm B} \underset{2\tilde k^{\rm Bb}_i}{\overset{\tilde k^{\rm Bf}_i}{
\rightleftarrows}}
{\rm B}_2\widetilde{\rm C}_i
\label{eq:P1}
\\
&{\rm B}_x\widetilde{\rm C}_i + {\rm A} \underset{\tilde k^{\rm Ab}_i}{\overset{x\tilde k^{\rm Af}_i}{
\rightleftarrows}}
{\rm A}{\rm B}_x\widetilde{\rm C}_i, 
{\rm A}{\rm B}_2\widetilde{\rm C}_i + {\rm A} 
\underset{2\tilde k_i^{\rm Ab}}{\overset{\tilde k_i^{\rm Af}}{\rightleftarrows}}
{\rm A}_2{\rm B}_2\widetilde{\rm C}_{i}
\label{eq:P2}
\\
&{\rm C}_i \underset{k_{\rm dps}}{\overset{k_{\rm ps}}{\rightleftarrows}} {\rm C}_{i+1
},
\widetilde{{\rm C}}_i
\underset{\tilde{k}_{\rm dps}}{\overset{\tilde{k}_{\rm ps}}{\rightleftarrows}}
\widetilde{{\rm C}}_{i+1}
\label{eq:P3}
\\
&{\rm B}_x\widetilde{{\rm C}}_i
\underset{\tilde{k}_{\rm dps}}{\overset{\tilde{k}_{\rm ps}}{\rightleftarrows}}
{\rm B}_x\widetilde{{\rm C}}_{i+1},
{\rm A}_y{\rm B}_x\widetilde{{\rm C}}_i
\underset{\tilde{k}_{\rm dps}}{\overset{\tilde{k}_{\rm ps}}{\rightleftarrows}}
{\rm A}_y{\rm B}_x\widetilde{{\rm C}}_{i+1} \; . \label{eq:P4}
\end{alignat} 

This model reproduces the phosphorylation behaviour of KaiC {\em in
  vitro} not only when all Kai proteins are present, but also when
KaiA and/or KaiB are absent \cite{Zon:2007ly}.  It moreover correctly predicted the experimentally observed disappearance of oscillations when the KaiA concentration is raised \cite{Rust:2007rb,Nakajima:2010jx}, a success that 
strongly supports the idea that KaiA sequestration is the primary
driver of synchronization.  Our model does not feature monomer
exchange between KaiC hexamers, an alternative means of
synchronisation \cite{Emberly:2006kr} that has been observed in experiments \cite{Kageyama:2006jp,Mori:2007pd}; we and
others find that monomer exchange is not critical for
stable oscillations \cite{Rust:2007rb,Zon:2007ly,Clodong:2007ff}. In
the \textit{Supporting Information} ({\em SI}), we show that similar results are obtained with a model that focuses on
the phosphorylation cycle of individual KaiC monomers
\cite{Nishiwaki:2007ao,Rust:2007rb} rather than of KaiC hexamers and that
includes a positive feedback loop.

\begin{figure}[t]
\center
\includegraphics[width=4cm]{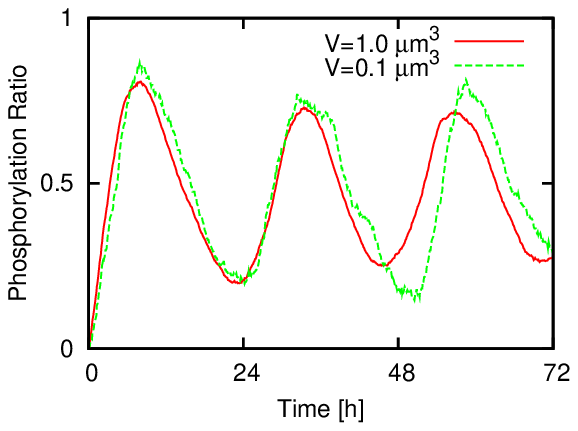}
\includegraphics[width=4cm]{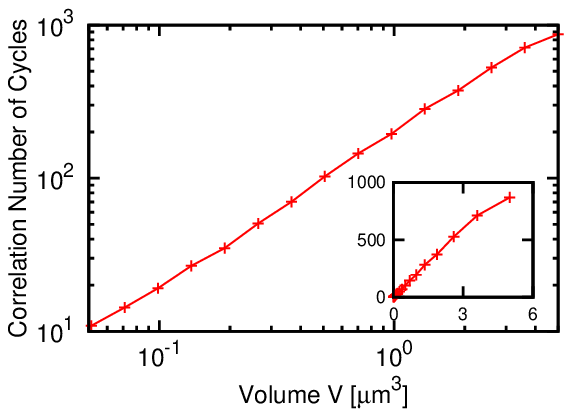}
\caption{The KaiC phosphorylation oscillator is highly stable when
  proteins are not synthesized or degraded.  (A) Time trace of the
  phosphorylation ratio $p(t)$, \textit{i.e.} the fraction of monomers
  that are phosphorylated, 
  for volumes $V=1 \mu {\rm m}^3$ and $V=0.1 \mu {\rm m}^3$. (B) Correlation
  number of cycles, $n_{1/2}$, as a function of the volume.  The
  protein concentrations are those used in the {\em in vitro}
  experiments \cite{Kageyama:2006jp,Nakajima:2010jx}: ${\rm
    [A]_T}=0.58\mu{M}; {\rm [B]_T} = 1.75\mu{\rm M}; {\rm [C]_T} =
  0.58 \mu{\rm M}$. For other parameters, see Table S1 of {\em
    SI}. \label{fig:PPC_Only}}
\end{figure}

We quantify our model's robustness to chemical noise by performing
kinetic Monte Carlo simulations of the chemical master equation
\cite{Gillespie77}. In our simulations, we vary the reaction volume,
but keep the concentrations of the Kai proteins constant at levels
comparable to those used in the {\em in vitro} experiments
\cite{Kageyama:2006jp,Nakajima:2010jx}.  Fig. \ref{fig:PPC_Only}A
shows a time trace of the fraction $p(t)$ of phosphorylated KaiC monomers
at a volume of order that of a cyanobacterium, while
Fig.~\ref{fig:PPC_Only}B gives, as a function of volume, the
correlation number of cycles $n_{1/2}$, defined as the
number of cycles after which the standard deviation in the phase of
the oscillation is half a day \cite{Eguchi:2008sw}; in the {\em SI} we
discuss our error estimates, which suggest that our computed $n_{1/2}$
values are typically accurate to within 15--20\%. One issue that
arises in comparing the simulation results to the measured
\textit{in vivo} clock robustness is that the Kai proteins appear to
be present in living cells in a ratio at which the \textit{in vitro}
system would not oscillate.  Kitayama \textit{et al.}
found that there are approximately 10,000 KaiC monomers but only 250--500
KaiA monomers per \textit{S. elongatus} cell~\cite{Kitayama:2003ts},
corresponding to a roughly 6:1 KaiC hexamer to KaiA dimer ratio
instead of the standard 1:1 ratio \textit{in
  vitro}~\cite{Kageyama:2006jp,Nakajima:2010jx}.  It has been
suggested that this discrepancy may indicate that the clock reactions
are in fact confined to a subdomain of the whole cell from which some
of the KaiB and KaiC molecules are excluded~\cite{Kitayama:2003ts};
here, we adopt this hypothesis and assume that the Kai proteins are
found in the physiologically relevant reaction volume in
proportions comparable to those used in the \textit{in vitro}
experiments.  If we
take this volume to be $\sim 1 \mu \text{m}^3$, comparable to the size of the
entire cell, then Fig.~\ref{fig:PPC_Only}B shows that $n_{1/2} \approx
200$, on the high side of the measured $166 \pm
100$~\cite{Mihalcescu:2004yq}.  Even if we consider a volume of only
$0.5 \mu \text{m}^3$---small enough that the measured number of KaiA
molecules is adequate to give the \textit{in vitro} KaiA dimer 
concentration of $0.58 \mu \text{M}$---we
find that $n_{1/2} \approx 102$, still within the experimental bounds
(in contrast to the predictions of some alternative
models~\cite{Rust:2007rb,Eguchi:2008sw}; see {\em SI}).  Our model thus predicts
that the PPC is very resistant to noise arising from the intrinsically
stochastic nature of chemical reactions.

\subsection{A phosphorylation cycle with constant protein synthesis and
  degradation rates is not stable}
Fig. \ref{fig:PPC_Only} shows that the phosphorylation cycle is highly
robust when the total concentrations of the Kai proteins are strictly constant.
But {\em in vivo} proteins are constantly being synthesized and
degraded. To study how this affects the PPC, we consider a model in which
the Kai proteins are produced and degraded in a stochastic
(memoryless) fashion with rates that are constant in time, with the
effects of active degradation~\cite{Imai:2004oc} and of passive
dilution lumped into a single first-order decay rate $\mu$ (see \textit{SI}).

\begin{figure}[t]
\center
\includegraphics[width=4cm]{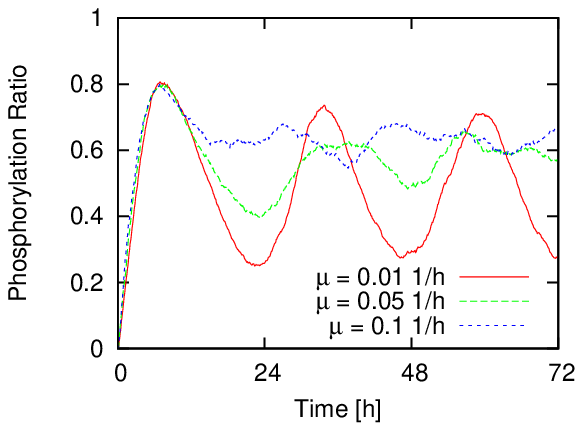}
\includegraphics[width=4cm]{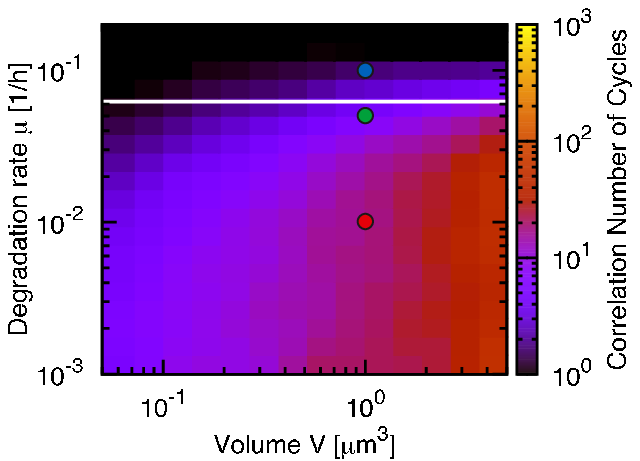}
\caption{A system in which the Kai proteins are produced and degraded
  with rates that are constant in time cannot sustain a stable
  phosphorylation cycle. (A) Time trace of the phosphorylation ratio
  $p(t)$ for three different degradation rates $\mu$ and $V=1\mu{\rm m}^3$. (B) Correlation
  number of cycles, $n_{1/2}$, as a function of the volume and
  degradation rate.  The white line denotes the bifurcation line $\mu = 0.0621 \mathrm{h}^{-1}$,
  where the system undergoes a supercritical Hopf bifurcation in the
  deterministic limit; the three colored dots correspond to the three
  curves in panel A. All
  proteins are degraded with the same rate; KaiA, KaiB and KaiC are produced with rates such that the average
  total concentrations equal those used in the \textit{in vitro}
  experiments (see Fig. \ref{fig:PPC_Only}). \label{fig:PPC_Prod}}
\end{figure}

Fig. \ref{fig:PPC_Prod} shows the performance of the phosphorylation
cycle as a function of degradation rate and cell volume; in
all cases, the synthesis rates are adjusted so that the mean
concentrations are constant and equal to those used in the previous section.
The oscillator's robustness clearly decreases dramatically
with increasing protein synthesis and decay rate. For a volume
comparable to that of a cyanobacterium and a degradation rate of $0.03
{\rm hr}^{-1}$, the correlation time is less than 20 days, much lower
than that observed {\em in vivo}
\cite{Mihalcescu:2004yq}. This degradation rate is precisely the
effective rate arising from protein dilution
with a cell doubling time of 24 hours. It is known, however, that KaiC
is also degraded actively at a rate as high as $0.1
{\rm hr}^{-1}$ \cite{Imai:2004oc}, leading to still worse stability.

The disappearance of the oscillations for higher protein synthesis and
decay rates can be understood by noting that fresh KaiC hexamers are
made in a fixed phosphorylation
state, which then has to catch up with those of the proteins
that are already in the cycle~\cite{Johnson:2008jy,Ito:2007ca}. When the degradation rate is high,
 the new proteins are likely to be degraded before the PPC can synchronise their phosphorylation levels;
indeed, in the limit that the protein synthesis and decay rates go to
infinity, $p(t)$ becomes constant in time and equal to the phosphorylation level of freshly made KaiC proteins. This is not a purely stochastic effect; the
dashed bifurcation line of Fig. \ref{fig:PPC_Prod}B shows that
 even in a deterministic model the oscillations disappear when the
synthesis and decay rates become too big (see \textit{SI}).  

\subsection{A protein phosphorylation cycle with a
  transcription-translation cycle is very stable}
To sustain a phosphorylation cycle, KaiC has to be made in an
oscillatory fashion: newly synthesized KaiC proteins have to be
injected into the phosphorylation cycle at the moment that the
phosphorylation states of the other proteins that are already in the
cycle match the modification state of the fresh KaiC proteins
\cite{Johnson:2008jy}. This is the principal role of the
transcription-translation cycle. Here, we present a model for how such
a cycle might interact with the protein phosphorylation cycle (see
Fig. \ref{fig:cartoon}). The
model is inspired by that of
Kondo {\em et al.} \cite{Taniguchi:2010tt} and contains the following
key ingredients:\\
\noindent {\em 1. RpaA activates \textrm{kaiBC} expression}\\
Deletion of {\em rpaA}
reduces the expression of clock-controlled genes, including that of
{\em kaiBC}
\cite{Taniguchi:2007pf,Taniguchi:2010tt,Takai:2006cz}.  Since neither promoters
nor transcription or chromosome-compaction factors have been
identified that interact with RpaA \cite{Takai:2006cz}, we
 make the phenomenological assumption
that RpaA directly activates {\em kaiBC} expression \cite{Taniguchi:2010tt}. \\
\noindent {\em 2. RpaA is activated by KaiC when KaiC is in the active
  state}\\
RpaA is activated via phosphorylation by the histidine kinase SasA, whose activity is in turn stimulated by KaiC
\cite{Smith:2006hf,Takai:2006cz}; inactivation of SasA reduces
{\em kaiBC} expression
\cite{Taniguchi:2007pf,Taniguchi:2010tt,Takai:2006cz}. Moreover,
SasA-RpaA phosphorylation occurs 4-8 hours before the peak of KaiC
phosphorylation \cite{Takai:2006cz}. This suggests that partially
phosphorylated KaiC that is on the active branch activates RpaA
through SasA \cite{Taniguchi:2010tt}.  Since SasA phosphorylation, occurring on time
scales of minutes \cite{Smith:2006hf}, is much faster than KaiC
phosphorylation, occurring on time scales of hours, we assume that the SasA dynamics can
be integrated out. \\
\noindent {\em 3. RpaA is inactivated by KaiC when KaiC is in the
  inactive state}\\
Inactivation of LabA \cite{Taniguchi:2007pf} or CikA \cite{Taniguchi:2010tt} increases {\em kaiBC} expression, with
inactivation of both having a still stronger effect
\cite{Taniguchi:2010tt}.  SasA inactivation can compensate for both
LabA \cite{Taniguchi:2007pf} and CikA inactivation
\cite{Taniguchi:2010tt}, but RpaA inactvation cannot compensate for
LabA inactivation \cite{Taniguchi:2007pf}. Taken together, these
results suggest that SasA, LabA, and CikA control {\em kaiBC}
expression through different pathways, with at least the SasA and LabA
pathways converging on RpaA
\cite{Taniguchi:2010tt}. Because phosphorylation of KaiC is critical
for negative feedback on {\em kaiBC} expression
\cite{Nishiwaki:2004kc}, LabA and CikA appear to act downstream of
phosphorylated KaiC \cite{Taniguchi:2010tt}. Since the mechanisms by
which LabA and/or CikA repress RpaA activation are unknown, we make
the phenomenological assumption
that KaiC on the inactive branch deactivates RpaA.\\
\noindent {\em 4. KaiC is injected into the system as fully
  phosphorylated hexamers} \\
Imai {\em et al.} reported that newly synthesized KaiC is
phosphorylated \textit{in vivo} within 30 min \cite{Imai:2004oc}, much
faster than phosphorylation of KaiC hexamers \textit{in vitro}, which
takes about 6 hours \cite{Tomita:2005bn};  
we thus assume that
newly synthesized KaiC is injected into the
system as fully phosphorylated KaiC hexamers. \\
\noindent {\em 5. KaiA and RpaA are synthesized at constant rates} \\
The mRNA levels of {\em kaiA} and {\em rpaA} exhibit much weaker oscillations than those of {\em kaiB} and {\em kaiC}
\cite{Ito:2009dm}. We therefore assume that {\em kaiA} and {\em rpaA}
are expressed at constant rates.\\
\noindent {\em 6. The phosphorylation cycle in vivo is similar to that
  in vitro}.\\

Fig. \ref{fig:cartoon} shows a cartoon of this model, which is
described by the reactions of Eqs. \ref{eq:P0}--\ref{eq:P4}
for the PPC together with the following reactions for the
TTC and the coupling between them:\\
\begin{alignat}{1}
  &\widetilde{\rm R} + {\rm X} \overset{k_{\rm a}}\rightarrow {\rm
    R} + {\rm X},\,\,\, {\rm R} + \widetilde{{\rm X}}
  \overset{k_{\rm i}} \rightarrow \widetilde{\rm R} + \widetilde{{\rm X}}  \label{eq:PT0}\\
  &\emptyset\underset{\tau;\sigma_\tau}{\overset{\beta_{\rm c}{\rm
        [R]}^4/(K^4+{ \rm [R]}^4)}\Longrightarrow}
  {\rm C}_6 + 3 {\rm B} \label{eq:PT1}\\
  &\emptyset \overset{\beta_{\rm a}} \rightarrow {\rm A},
  \,\,\,\emptyset
  \overset{\beta_{\rm r}} \rightarrow {\rm \widetilde{R}}, \label{eq:PT2}\\
  &{\rm R},\widetilde{\rm R},{\rm A}, {\rm B},{\rm
    A}{\rm C}_i,{\rm B}_x\widetilde{{\rm C}}_i,{\rm A}_y{\rm B}_x\widetilde{{\rm C}}_i
  \overset{\mu}\rightarrow \emptyset
\label{eq:PT3}
\end{alignat}
Eq. \ref{eq:PT0} models activation of inactive RpaA, $\widetilde{\rm
  R}$, by ${\rm X}\in\{{\rm A}{\rm C}_2,\dots {\rm A}{\rm C}_5\}$ and
inactivation of active RpaA, R, by $\widetilde{{\rm X}} \in\{ {\rm
  A}_y {\rm B}_x\widetilde{{\rm C}}_5 \dots {\rm A}_y {\rm
  B}_x\widetilde{\rm C}_2\}$.  The model is robust to the precise
choice of phosphoforms that activate and repress RpaA (see {\em
  SI}). Because KaiA enhances {\em kaiBC} expression
\cite{Iwasaki:2002um}, we assume that active KaiC bound to KaiA
activates RpaA; a model in which both ${\rm C}_i$ and ${\rm A}{\rm
  C}_i$ activate RpaA also generates oscillations, albeit less
robustly (see {\em SI}).  Eq. \ref{eq:PT1} models the activation of
{\em kaiBC} expression by RpaA, using a Hill function with coefficient
4. We assume a normally distributed delay, denoted by
the double arrow, with mean $\tau = 5$ hr and standard deviation
$\sigma_\tau = 0.5$ hr, between the activation of \textit{kaiBC}
transcription and the appearance of KaiB and KaiC protein.  The length
of the delay is dictated by the requirement that new KaiC protein be
produced when its phosphorylation state matches that of the PPC.
Eq. \ref{eq:PT2} models constitutive expression of \textit{kaiA} and
\textit{rpaA}. Eq. \ref{eq:PT3} describes degradation of all species
with the same rate constant $\mu$; we ignore rhythmic KaiC
degradation~\cite{Imai:2004oc}, which is not essential to produce a
robust clock. In the {\em SI} we discuss a more detailed model
that includes cell growth and binomial partitioning upon cell
division, which appreciably increases noise without changing
qualitative trends. Although one can think of our model in loose terms as
consisting of coupled transcriptional and post-translational
oscillators, it cannot be mathematically decomposed into two separate
oscillatory systems, each with its own variables, nor can the strength
of the coupling between the two cycles be independently tuned.  It is
thus formally quite different from textbook models of coupled
oscillators ~\cite{Syncbook}.

\begin{figure}[t]
\center
\includegraphics[width=4cm]{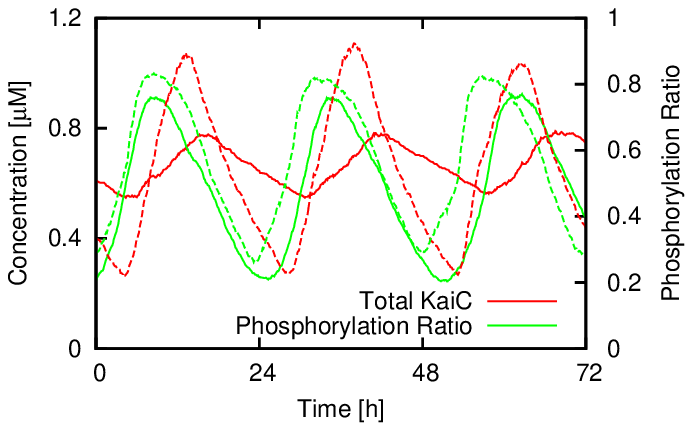}
\includegraphics[width=4cm]{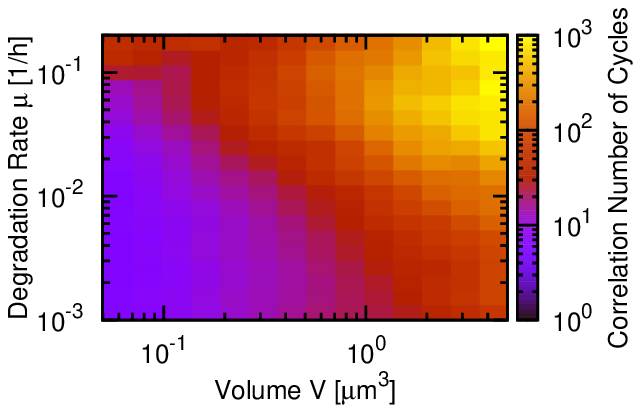}
\caption{The combination of a TTC and a PPC can generate stable
  circadian rhythms. (A) Time trace of the phosphorylation ratio
  $p(t)$ and total KaiC concentration ${\rm [C]_T}$ for $V=1\mu{\rm m}^3$
  and $\mu=0.03 {\rm hr}^{-1}$ (solid lines) and $0.1 {\rm hr}^{-1}$
  (dashed lines). (B) Correlation number of cycles,
  $n_{1/2}$, for $p(t)$ as a function of the volume and degradation rate
  $\mu$. The protein synthesis rates are varied with the degradation
  rates such that the average protein concentrations equal those used
  {\em in vitro} (see Fig. \ref{fig:PPC_Only}).\label{fig:PPC_TTC}}
\end{figure}

Fig. \ref{fig:PPC_TTC}B shows the robustness of this PPC-TTC model as
a function of cell volume and protein degradation rate $\mu$; when
$\mu$ is varied, all three protein synthesis rates $\beta_{\alpha}$
are adjusted so that the average protein concentrations are unchanged
and comparable to those of the models considered above.  As expected,
$n_{1/2}$ decreases with decreasing cell volume. Its dependence on the
degradation rate, however, is markedly different from that seen with
constant KaiC synthesis (Fig.~\ref{fig:PPC_Prod}B): A PPC sustained by
a TTC becomes \textit{more} robust with increasing decay rate. If we
assume that proteins are lost only through dilution, then for a
bacterial volume of $1\,\mu\mathrm{m}^3$ and a cell doubling time of
24 hours (corresponding to a decay rate $\mu = 0.03 {\rm hr}^{-1}$),
the correlation time is about 200 days, consistent with the value
measured experimentally \cite{Mihalcescu:2004yq}.  If proteins are
also degraded actively, increasing $\mu$, this excellent behavior
improves still further; even for $V=0.5\mu{\rm m}^3$, $n_{1/2}\approx
120$ for $\mu=0.1 {\rm hr}^{-1}$.  This is in stark contrast to the
stability of a PPC without a TTC, which we have seen is hardly stable
for a protein lifetime of about 10 hours (Fig. \ref{fig:PPC_Prod}B).

\subsection{A protein phosphorylation cycle dramatically enhances the
  robustness of a transcription-translation cycle}
Figs. \ref{fig:PPC_Prod}B and \ref{fig:PPC_TTC}B show that a TTC can
greatly improve the robustness of a PPC.  One might thus ask whether
the PPC is needed at all, or whether an adequate clock can be built
with only a TTC.  To address this question, we modify the PPC-TTC model (Eqs. \ref{eq:PT0}--\ref{eq:PT3})
so that it consists only of a TTC:
\begin{alignat}{1}
  &\emptyset\underset{\tau;\sigma_\tau}{\overset{\beta_{\rm c}{\rm
        [R]}^4/(K^4+{ \rm [R]}^4)}\Longrightarrow}
  {\rm C},\,\,\,\emptyset \overset{\beta_{\rm r}}\rightarrow
  {\rm R},\,\,\,{\rm C}, {\rm R} \overset{\mu}\rightarrow \emptyset\label{eq:T1}\\
  &\widetilde{\rm R}\overset{k_{\rm a}^{\rm t}}\rightarrow {\rm R},\,\,\,{\rm C} +
    {\rm R} \overset{k_{\rm i}^{\rm t}}\rightarrow {\rm C} + \widetilde{\rm R}
\end{alignat}
In the simulations, we adjust the
delay $\tau$ and the synthesis rate $\beta$ for each choice of the degradation rate $\mu$ such that the oscillation
period is 24 hours and the average KaiC
concentration is comparable to that of the other models considered so
far (see {\em SI}). We also investigated the simplest possible TTC
model, namely one in which KaiC directly represses its own synthesis; this
gave very similar results (see {\em SI}).

\begin{figure}[t]
\center
\includegraphics[width=4cm]{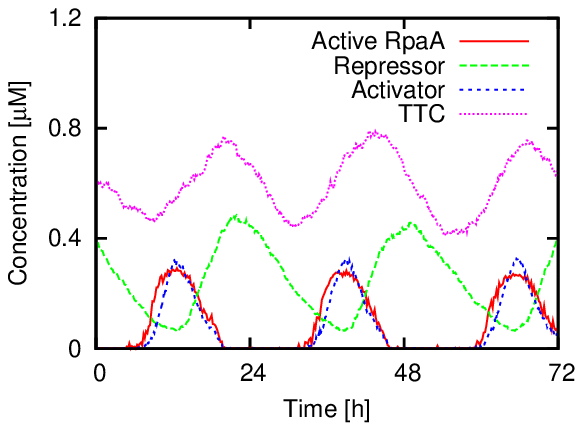}
\includegraphics[width=4cm]{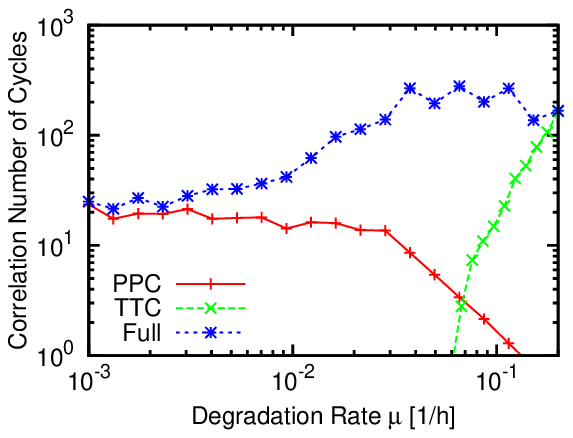}
\caption{A PPC in combination with a TTC generates robust rhythms over
  a wide range of degradation rates. (A) Time traces of the concentrations of active RpaA and of the KaiC phosphoforms that activate (``activator'') and repress
  (``repressor'') RpaA for the PPC-TTC model ($V=1 \mu{\rm m}^{3}$
  and $\mu=0.03{\rm hr}^{-1}$), and KaiC for the
  TTC-only model ($V=1\mu{\rm m}^3$ and $\mu=0.12 {\rm hr}^{-1}$; for
  $\mu=0.03 {\rm hr}^{-1}$ the TTC does not oscillate); (B)
  Correlation number of cycles as a function of degradation rate $\mu$
  for the PPC, TTC, and combined TTC-PPC model  ($V=1 \mu m^3$); for
  the TTC, $n_{1/2}$ was computed for ${\rm [C]}(t)$ rather than for $p(t)$.  The
  protein synthesis rates are varied with the degradation rates such
  that the average concentrations equal those used in the {\em in
    vitro} experiments (see Fig. \ref{fig:PPC_Only}).
  \label{fig:PPC_TTC_FULL}}
\end{figure}

Fig. \ref{fig:PPC_TTC_FULL}B shows for a bacterial volume of
approximately $1\,\mu\mathrm{m}^3$ the robustness of this TTC-only model as
a function of the degradation rate, together with the results for the
PPC-TTC model (Fig. \ref{fig:PPC_TTC}B) and for the PPC-only model with constant synthesis rates (Fig. \ref{fig:PPC_Prod}B).  The TTC-only model's behavior is the opposite of that of the PPC-only model; the TTC-only model is most stable for {\em high} degradation rates, and its robustness falls dramatically
when $\mu$ drops below $0.2 {\rm hr}^{-1}$. The
combined TTC-PPC model, however, is robust for all degradation
rates---its correlation time interpolates between those of the TTC-only model
for large $\mu$ and of the PPC-only model for small $\mu$. Importantly, the relative advantage of the combined model is
greatest when $1/\mu$ is of order the oscillator
period, and this is precisely the regime where physiological
degradation and dilution rates for \textit{S. elongatus}
fall~\cite{Imai:2004oc}.  Further, the combined oscillator does much
better than either oscillator alone; in contrast, when two
conventional phase oscillators with comparable noise levels are
coupled, one expects only about a factor of two gain in
$n_{1/2}$~\cite{Syncbook}.

Dilution puts a lower bound on the degradation rate, which means that
a stable oscillator cannot be based on a PPC only, especially when the
growth rate of the bacterium is large. The degradation rate can,
on the other hand, be increased by active degradation. For high degradation
rates, {\em i.e.} $\mu > 0.2{\rm hr}^{-1}$, an oscillator based
only on transcriptional feedback can be sufficiently robust
(Fig. \ref{fig:PPC_TTC_FULL}B). However, to balance these high decay
rates, the protein synthesis rates have to be correspondingly large,
which is energetically costly \cite{Dekel:2005fj}.
Combining a PPC with a TTC makes it possible to dramatically improve
the robustness while keeping the protein synthesis rates the
same. While a phosphorylation cycle does not come entirely for
free \cite{Terauchi:2007hk},
this suggests that a PPC combined with a TTC
gives the best performance-to-cost ratio.

The question remains why a PPC so effectively enhances the robustness
of a TTC. Fig. \ref{fig:PPC_TTC_FULL}A shows, for the full PPC-TTC
model, time traces of the concentration of RpaA, the sum of the
concentrations of the KaiC phosphoforms that activate RpaA, and the sum of those
that repress RpaA; for comparison, we also show the
KaiC concentration in the TTC-only model. In the TTC-only model, the KaiC concentration varies slowly, and
the concentration threshold for {\em kaiBC} repression is crossed at a
shallow angle. In contrast, in the PPC-TTC model the concentration of
active RpaA switches rapidly between a value that is well above the
activation threshold for {\em kaiBC} expression and one that is well
below it.  Such sharp threshold crossings minimize the effect of fluctuations in the concentration of the
regulatory protein on the timing of the switch between {\em kaiBC}
activation and {\em kaiBC} repression. RpaA's abrupt switching is a direct consequence of the sharp rise and fall in the concentrations of the KaiC phosphoforms that regulate RpaA. This in turn stems not so much from the oscillation in the
total concentration of KaiC, but from the phosphorylation cycle
at the multiple phosphorylation sites of KaiC:  The concentrations of
the various phosphoforms approach zero in certain parts of the cycle
(Fig. \ref{fig:PPC_TTC_FULL}A), even though the total KaiC
concentration remains relatively large throughout
(Fig. \ref{fig:PPC_TTC}A). Indeed, the stability of the circadian
rhythm in the PPC-TTC model is mostly determined by the KaiC
phosphorylation cycle, which, as Fig. \ref{fig:PPC_Only}B shows, is
highly robust. It thus seems that the phosphorylation cycle is the
principal pacemaker, even though a transcription-translation cycle is
needed to sustain it and can, in fact, raise its robustness above the
limiting value at zero protein synthesis and decay rate
(Fig. \ref{fig:PPC_TTC_FULL}B).

\section{Discussion}
The evidence is accumulating that circadian rhythms are driven by both
transcription-translation and protein-modification cycles not just in
cyanobacteria but even in higher
organisms~\cite{Johnson:2008jy,Merrow:2007zr}.  Our analysis suggests
that both cycles are required to generate stable circadian rhythms in
growing cells over a wide range of conditions.  On the one hand, it is
intrinsically difficult for a TTC to generate oscillations when the
average protein decay time is long compared to the oscillation period.
Indeed, the speed of protein degradation (and/or dilution) sets a
simple upper bound on the size of any oscillation; if only a small
fraction of the proteins can be degraded in one oscillation cycle,
then any modulation of the total protein concentration must
necessarily have a low amplitude and so be very susceptible to noise.
In cells whose doubling time is of order tens of hours (comparable to
the 24 hour clock period) or longer, a TTC alone can thus function
robustly only at the expense of high rates of protein synthesis and
active protein degradation.  On the other hand, 
a protein modification cycle must fail when protein turnover becomes
faster than the clock period: Proteins are then typically degraded and
replaced by new molecules in the wrong modification state before they
can pass through the full cycle, destroying the
oscillation.  While a TTC is thus required to sustain a protein
modification cycle at higher growth rates, a modification cycle can
also greatly enhance the performance of a TTC, especially at
lower protein synthesis and decay rates, where the
modification cycle can be exploited to set the rhythm of gene
expression.
The protein phosphorylation cycle of KaiC is intrinsically very
robust, because KaiC is present in relatively large copy numbers and because a
KaiC hexamer must go through many individual phosphorylation and
dephosphorylation steps in each oscillation period.  These two
characteristics together are responsible for the reliable and
pronounced oscillations in the concentrations of the various
phosphoforms of KaiC (Fig. \ref{fig:PPC_TTC_FULL}A), which, in turn,
drive the robust rhythm of {\em kaiBC} expression.

Our combined PPC-TTC model is consistent with a number of experimental
observations. It not only matches the average oscillations of the
phosphorylation level and the total KaiC concentration in wildtype
cells (Fig. \ref{fig:PPC_TTC}A), but also reproduces the observation
that even in the presence of an excess of KaiA, the total KaiC
concentration still oscillates with a circadian period
\cite{Kitayama:2008kz} (see {\em SI}).  Yet, because quite a few
elements of the TTC have not been characterized experimentally, our
model of the TTC is necessarily rather simplified and
phenomenological; not surprisingly, some observations thus cannot be
reproduced. For example, our model predicts that the phase of the
oscillation in KaiC abundance lags behind that of the phosphorylation
level by a few hours, while experiments seem to show that these
oscillations are more in phase \cite{Tomita:2005bn,Kitayama:2008kz}. This may be due
to our simplifying assumption that KaiC is produced as fully
phosphorylated hexamers. While phosphorylation of fresh KaiC has been
reported to occur within 30 minutes \cite{Imai:2004oc}, fragmentary
evidence suggests that hexamerisation is slow~\cite{Mori:2002sy}; one
would then expect to detect KaiC monomers in experiments several hours
before our simulations report the presence of fully functional
hexamers.   We believe, however,
that while our model is simplified, it does capture the essence of a
coupled TTC and PPC: that is, it is built around a protein that not
only regulates its own synthesis, but also undergoes a protein
modification cycle, where the different modification states either
activate or repress protein synthesis.  Moreover, the basic trends we
observe can largely be explained by the arguments of the preceding
paragraph, which depend only on the comparison of a protein decay
timescale with the oscillation period; we thus expect that our
qualitative results should apply to any system that exploits both a
protein modification cycle and a transcription-translation cycle to
generate circadian rhythms.

In {\em S. elongatus}, the period of the circadian rhythm is
insensitive to changes in the growth rate
\cite{Mori:1996pa,Kondo:1997bs}. Since protein synthesis and decay
rates, on the other hand, tend to vary with the growth rate \cite{Klumpp:2009uj}, the
question arises whether the period of the oscillator as predicted by
our model is robust to such variations. In the {\em SI} we show that a
ten-fold increase in the synthesis and degradation rates of all
proteins only decreases the period by 10-20\%. The reason is that the
period of the oscillator is mostly determined by the intrinsic period
of the phosphorylation cycle (Fig. \ref{fig:PPC_TTC_FULL}A). The
latter does not depend on the absolute protein synthesis and decay
rates, although it does depend on the ratio of the concentrations of
the Kai proteins
\cite{Kageyama:2006jp,Nakajima:2010jx,Zon:2007ly}. Clearly, it would
be of interest to investigate how the ratio of the concentrations of
the Kai proteins varies with the growh rate.

Finally, how could our predictions be tested experimentally? The most
important testable prediction of our analysis is that the PPC requires
a TTC when the growth rate is high. Kondo and coworkers have
demonstrated that a PPC can function in the absence of a TTC
\cite{Tomita:2005bn}. These experiments were performed under constant
dark (DD) conditions, which means that the growth rate was
(vanishingly) small, protein synthesis was halted,
and even protein decay was probably negligible, since the KaiC level
was rather constant \cite{Tomita:2005bn}. These experiments are
consistent with our predictions, but the critical test would be to increase the growth rate while avoiding any circadian regulation of \textit{kaiBC} expression. This means that the cyanobacteria must
be grown under constant light (LL) conditions. Moreover, one would
need to bring {\em kaiBC} expression under the control of a promoter
that is constitutively active. However, most promoters of {\em
  S. elongatus} \cite{Ito:2009dm,Vijayan:2009rq}, and even many heterologous promoters
from {\em E. coli} \cite{Ditty:2003vy}, are influenced by the circadian clock.
 Nevertheless, a number of
promoters have been reported that exhibit arhythmic activity
\cite{Ito:2009dm}, and these might be possible candidates. A
still more challenging experiment would be to express the phosphorylation
cycle in the bacterium {\em E. coli}~\cite{Tozaki:2008px}. 
Our analysis predicts that in
growth-arrested cells, the phosphorylation cycle should be functional,
while in normal growing {\em E. coli} cells, with cell doubling times
of roughly 1 hour, the oscillations should cease to exist.

\subsection{Acknowledgments}
We thank Tom Shimizu for critically reading the manuscript. This
work was supported in part by FOM/NWO (D.Z. and P.R.t.W.) and by NSF grant PHY05-51164 to the KITP (D.K.L.).


\onecolumngrid
\newpage

\newcommand{\abs}[1]{\left| #1 \right|}
\newcommand{\mean}[1]{\left\langle #1 \right\rangle}
\newcommand{\wiener}{\mathcal{W}}

\renewcommand{\thefigure}{S\arabic{figure}}
\renewcommand{\theequation}{S\arabic{equation}}
\renewcommand{\thetable}{S\arabic{table}}

\setcounter{section}{0}
\setcounter{figure}{0}
\setcounter{equation}{0}
\setcounter{page}{0}
\setcounter{table}{0}

\begin{center}
{\bf \Large Supporting Information}\\[0.2cm]
{\it \large Robust circadian clocks from coupled protein modification and
  transcription-translation cycles}\\[0.2cm]
{\large David Zwicker, David K. Lubensky and Pieter Rein ten Wolde}
\end{center}

\section{The model}
\label{sec:Model}
\subsection{Protein phosphorylation cycle}
The model for the KaiC protein phosphorylation cycle (PPC) is
described by reactions (1)-(5) of the main text, which we repeat here:
\begin{align}
&{\rm C}_i \underset{b_i}{\overset{f_i}{\rightleftarrows}}
  \widetilde{{\rm C}}_i,
&&{\rm C}_i + {\rm A} 
\underset{k_i^{\rm Ab}}{\overset{k_i^{\rm Af}}{\rightleftarrows}}
{\rm AC}_i
  \overset{k_{\rm pf}}{\rightarrow} {\rm C}_{i+1} + {\rm A},
\label{eqS:P0}
\\
&\widetilde{\rm C}_i + {\rm B} \underset{\tilde k^{\rm Bb}_i}{\overset{2\tilde k^{\rm Bf}_i}{
\rightleftarrows}}
{\rm B}\widetilde{\rm C}_i, 
&&
{\rm B}\widetilde{\rm C}_i + {\rm B} \underset{2\tilde k^{\rm Bb}_i}{\overset{\tilde k^{\rm Bf}_i}{
\rightleftarrows}}
{\rm B}_2\widetilde{\rm C}_i,
\label{eqS:P1}
\\
&{\rm B}_x\widetilde{\rm C}_i + {\rm A} \underset{\tilde k^{\rm Ab}_i}{\overset{x\tilde k^{\rm Af}_i}{
\rightleftarrows}}
{\rm A}{\rm B}_x\widetilde{\rm C}_i, 
&&
{\rm A}{\rm B}_2\widetilde{\rm C}_i + {\rm A} 
\underset{2\tilde k_i^{\rm Ab}}{\overset{\tilde k_i^{\rm Af}}{\rightleftarrows}}
{\rm A}_2{\rm B}_2\widetilde{\rm C}_{i},
\label{eqS:P2}
\\
&{\rm C}_i \underset{k_{\rm dps}}{\overset{k_{\rm ps}}{\rightleftarrows}} {\rm C}_{i+1
},
&&
\widetilde{{\rm C}}_i
\underset{\tilde{k}_{\rm dps}}{\overset{\tilde{k}_{\rm ps}}{\rightleftarrows}}
\widetilde{{\rm C}}_{i+1},
\label{eqS:P3}
\\
&{\rm B}_x\widetilde{{\rm C}}_i
\underset{\tilde{k}_{\rm dps}}{\overset{\tilde{k}_{\rm ps}}{\rightleftarrows}}
{\rm B}_x\widetilde{{\rm C}}_{i+1},
&&
{\rm A}_y{\rm B}_x\widetilde{{\rm C}}_i
\underset{\tilde{k}_{\rm dps}}{\overset{\tilde{k}_{\rm ps}}{\rightleftarrows}}
{\rm A}_y{\rm B}_x\widetilde{{\rm C}}_{i+1} \; . \label{eqS:P4}
\end{align}
Here, ${\rm C}_i$ denotes a KaiC hexamer in the active conformational
state, in which the number $i$ of phosphorylated monomers tends to
increase, and ${\rm \widetilde{C}}_i$ denotes a KaiC hexamer in the
inactive conformational state in which $i$ tends to decrease; ${\rm
  A}$ denotes a KaiA dimer, and ${\rm B}$ denotes a KaiB dimer.  The
reactions ${\rm C}_i \rightleftarrows \widetilde{{\rm C}}_i$ in
Eq. \ref{eqS:P0} model the conformational transitions between active
and inactive KaiC; the second set of reactions in Eq. \ref{eqS:P0}
describe phosphorylation of active KaiC that is stimulated by KaiA; the
reactions in Eqs. \ref{eqS:P1} and \ref{eqS:P2} model the sequestration
of KaiA by inactive KaiC that is bound to KaiB; note that an inactive
KaiC hexamer can bind up to two KaiA dimers; the reactions in
Eqs. \ref{eqS:P3} and \ref{eqS:P4} model spontaneous phosphorylation
and dephosphorylation of active and inactive KaiC. For a more detailed
discussion of the model, we refer to \cite{Zon:2007ly}.

\subsection{A PPC with constant protein synthesis and degradation
  rates}
\label{sect:PPC-const-degrad}
In the main text, we also discuss the performance of a model that
includes not only a PPC, but also synthesis and degradation of the Kai
proteins. These reactions are given by
\begin{align}
&\emptyset \overset{\beta_{\rm c}}\rightarrow {\rm C}_6 + 3 {\rm
  B},&\emptyset \overset{\beta_{\rm a}}\rightarrow {\rm A},\\
&{\rm A}, {\rm B},{\rm A}{\rm C}_i,{\rm B}_x\widetilde{{\rm C}}_i,{\rm A}_y{\rm B}_x\widetilde{{\rm C}}_i
  \overset{\mu}\rightarrow \emptyset \label{eqS:P5}
\end{align}
As explained in the main text, we assume that fresh KaiC is injected
into the system as fully phosphorylated hexamers because
phosphorylation of fresh KaiC proteins has been reported to be fast,
{\em i.e.}  occurring within 30 minutes \cite{Imai:2004oc}. However,
the precise choice for the phosphoform of fresh KaiC is not so
important in this model; it does not affect the robustness of this
model.  Because the KaiB and KaiC proteins are both products of the \textit{kaiBC} operon, we choose to model the production of both proteins as a single reaction.
 We note that while in the model with the
transcription-translation cycle, discussed in section \ref{sec:TTC_PPC}, the
delay in the synthesis reactions is critical, in the above model,
where the Kai proteins are produced with rates that are constant in
time, a delay would have no effect; the synthesis reactions are
therefore modeled as simple, Poissonian birth reactions.

\subsection{Deterministic PPC model with synthesis and degradation}
To verify that the disappearance of oscillations as the degradation
rate is increased is not a purely stochastic effect, we consider the
model of Eqs.~\eqref{eqS:P0}--\eqref{eqS:P5} in the deterministic limit of
infinite volume and protein number.  In this limit, the concentrations
of the different proteins evolve according to deterministic rate
equations.  We make two further simplifying assumptions:  First, we replace the two-step binding of KaiB to $\widetilde{\rm C}_i$ with a trimolecular reaction that turns $\widetilde{\rm C}_i$ directly into ${\rm B}_2\widetilde{{\rm C}}_i$, and making a similar change for binding of KaiA to the inactive branch.  Second, we assume that binding and unbinding reactions are
fast enough that they are effectively in steady state and thus
explicitly keep track only of the concentrations of the various KaiC
species; the concentrations of free KaiA and KaiB can then be inferred
from conservation laws.  The dynamical equations are then essentially
the same as those given in Eqs. 44 -- 47 of our previous
publication~\cite{Zon:2007ly},
with the addition of a linear decay term with rate $\mu$ for each species and of synthesis of $\mathrm{C}_6$ with rate $\beta_c$:
\begin{eqnarray}
\label{eqS:reducemodel1}
\frac{d[{\rm C}_i]_{\rm T}}{dt} &=&     \sigma^{{\rm
                        ps}}_{i\!-\!1}\![{\rm C}_{i\!-\!1}]_{\rm T}
                        \!+\!\sigma^{{\rm dps}}_{i\!+\!1} [{\rm
                        C}_{i\!+\!1}]_{\rm T}
                        \!-\!(\sigma^{{\rm ps}}_i\!+\!\sigma^{\rm dps}_i)
                        [{\rm C}_i]_{\rm T}
                        \!-\!\sigma^{\rm Ff}_i [{\rm C}_i]_{\rm T}
                        \!+\!\sigma^{\rm Fb}_i [\widetilde{\rm
                        C}_i] \\
&& + \beta_c \delta_{i,6} - \mu [{\rm C}_i]_{\rm T}\\
\frac{d[\widetilde{\rm C}_i]}{dt} 
                &=&     \tilde{k}_{{\rm ps}} [\widetilde{\rm C}_{i\!-\!1}]
                        \!+\!\tilde{k}_{{\rm dps}} [\widetilde{\rm C}_{i\!+\!1}]
                        \!-\!(\tilde{k}_{{\rm ps}}\!+\!\tilde{k}_{{\rm dps}})
                        [\widetilde{\rm C}_i]
                        \!+\!\sigma^{\rm Ff}_i [{\rm C}_i]_{\rm T}
                        \!-\!\sigma^{\rm Fb}_i [\widetilde{\rm C}_i] \nonumber\\
                &&      - \tilde{\kappa}^{\rm Bf}_i([{\rm B]_T} - 2
                        \textstyle\sum_i \displaystyle
                        [{\rm B}_2\widetilde{\rm C}_i]_{\rm T})^2
                        [\widetilde{\rm C}_i]
                        + \frac{\tilde{\kappa}^{\rm Bb}_i \widetilde{\rm K}_i 
                        [{\rm B}_2\widetilde{\rm C}_i]_{\rm T}}
                        {\widetilde{\rm K}_i+[{\rm A}]^2} - \mu [\widetilde{\rm C}_i]\\
\frac{d[{\rm B}_2\widetilde{\rm C}_i]_{\rm T}}{dt} 
                &=&     \tilde{k}_{{\rm ps}} [{\rm B}_2\widetilde{\rm
                        C}_{i\!-\!1}]_{\rm T}
                        \!+\!\tilde{k}_{{\rm dps}} 
                        [{\rm B}_2\widetilde{\rm C}_{\!+\!1}]_{\rm T}
                        \!-\!(\tilde{k}_{{\rm ps}}\!+\!\tilde{k}_{{\rm dps}})
                        [{\rm B}_2\widetilde{\rm C}_i]_{\rm T}                         \nonumber \\
                 &&     + \tilde{\kappa}^{\rm Bf}_i({\rm [B]_T} - 2
                        \textstyle\sum_i \displaystyle
                        [{\rm B}_2\widetilde{\rm C}_i]_{\rm T})^2
                        [\widetilde{\rm C}_i]
                        - \frac{\tilde{\kappa}^{\rm Bb}_i \widetilde{\rm K}_i 
                        [{\rm B}_2\widetilde{\rm C}_i]_{\rm T}}
                        {\widetilde{\rm K}_i+[{\rm A}]^2} - \mu [{\rm B}_2\widetilde{\rm C}_i]_{\rm T} \; ,
\label{eqS:reducemodel3}
\end{eqnarray}
where the concentration of free KaiA, [A], is given by
\begin{equation}
[{\rm A}]+\sum_{i=0}^N \frac{[{\rm A][C}_i]_{\rm T}}{{\rm K}_{i}+[{\rm A}]} 
+ 2 \sum_{i=0}^{N} \frac{[{\rm A}]^2[{\rm B}_2\widetilde{\rm
      C}_i]_{\rm T}}
{\widetilde{\rm K}_{i}^2+[{\rm A}]^2} - {\rm [A]_T} = 0 \label{eqS:Aeq} \; .
\end{equation}
Here $[{\rm C}_i]_{\rm T}$ is the total concentration of KaiC hexamers with $i$ phosphorylated monomers in the active state, whether or not complexed with KaiA, i.e $[{\rm C}_i]_{\rm T} = [{\rm C}_i] + [{\rm AC}_i]$; $[{\rm B}_2\widetilde{\rm C}_i]_{\rm T}$ is defined similarly.  The effective rate constants appearing in these equations depend on the concentration [A] of free KaiA and are defined in terms of the more microscopic rate constants as follows:  The effective (de)phosphorylation rates on the active branch
are $\sigma^{{\rm ps}}_i=(k_{\rm ps} {\rm K}_i+k_{\rm pf}[{\rm A}])/({\rm K}_i+[{\rm A}])$ and $\sigma^{{\rm dps}}_i={\rm K}_i k_{{\rm dps}}/({\rm K}_i+[{\rm A}])$.  The effective flipping
rates are given by $\sigma^{Ff}_i=f_iK_i/(K_i+[A])$ and $\sigma^{Fb}_i=b_i$, where $f_i$ and $b_i$ are the forward and backward flipping rates.  The parameters $\tilde{\kappa}^{{\rm Bf}}_{i}$ and $\tilde{\kappa}^{\rm Bb}_i$ differ from $\tilde{k}^{{\rm Bf}}_{i}$ and $\tilde{k}^{\rm Bb}_i$, respectively, in that the $\kappa$'s are rate constants for trimolecular reactions which are broken down into two successive bimolecular reactions in the stochastic simulations.  The dissociation constants ${\rm K}_i$  satisfy ${\rm K}_i=k^{{\rm Ab}}_{i}/k^{{\rm Af}}_{i}$; the $\widetilde{\rm K}_i$ could be defined similarly in terms of forward and backward rates for KaiA binding to the inactive branch, but (just as with $\tilde{\kappa}^{{\rm Bf}}_{i}$ and $\tilde{\kappa}^{\rm Bb}_i$) these rates would differ from those used in the stochastic simulations, so we choose instead to quote the dissociation constants directly.  Following~\cite{Zon:2007ly}, we choose values for the new parameters associated with trimolecular interactions such that time dependence of the phosphorylation ratio $p(t)$ matches the average behavior of the stochastic model.

To determine where oscillations disappear as $\mu$ is increased, we analyzed these equations using the XPPAUT implementation of the AUTO continuation package~\cite{Xppbook}.
We found that, for the parameter values given in Table~\ref{tbl:parameters}, the system undergoes a supercritical Hopf bifurcation at $\mu = \unitfrac[0.0621]{1}{h}$, as noted in the main text.

\subsection{The PPC and TTC combined}
\label{sec:TTC_PPC}
The full model of the main text consists of a PPC, a
transcription-translation cycle (TTC), and a
pathway that couples these two cycles. This model is described by the
reactions of Eqs. \ref{eqS:P0}-\ref{eqS:P4} for the PPC, together with the following
reactions for the TTC and the coupling between them:
\begin{align}
  &\widetilde{\rm R} + {\rm X} \overset{k_{\rm  a}}\rightarrow {\rm R} + {\rm X},&&
  {\rm R} + \widetilde{{\rm X}} \overset{k_{\rm i}} \rightarrow \widetilde{\rm R} + \widetilde{{\rm X}}, &
  \label{eqS:PT0}
\\
  &\emptyset
  \underset{ \tau \pm \sigma_{\tau}}{\overset{\beta_{\rm c}{\rm [R]}^4/(K^4+{\rm [R]}^4)}{\Longrightarrow}}
  {\rm C}_6 + 3 {\rm B},&& \emptyset \overset{\beta_{\rm a}} \rightarrow {\rm A},  &\emptyset
  \overset{\beta_{\rm r}} \rightarrow {\rm \widetilde{R}}, \label{eqS:PT1}
\\
&{\rm R},\widetilde{\rm R},{\rm A}, {\rm B},{\rm
    A}{\rm C}_i,{\rm B}_x\widetilde{{\rm C}}_i,{\rm A}_y{\rm B}_x\widetilde{{\rm C}}_i
  \overset{\mu}\rightarrow \emptyset \; .  \label{eqS:PT2}
\end{align}
Here, R and $\rm \widetilde R$ denote the RpaA protein in its active
and its inactive form, respectively. The ${\rm X}$ and
$\widetilde{{\rm X}}$ in Eq.~\eqref{eqS:PT0} denote any of the
phosphoforms of KaiC that mediate the activation and repression of
RpaA, respectively; in the next section, we discuss the dependence of
the results on precisely which phosphoforms are chosen. The double
arrow indicates a reaction with a Gaussian distributed delay with mean
$\tau$ and variance $\sigma_\tau$. We thus assume that {\em kaiBC}
expression is activated by RpaA, where the activity of RpaA is
modulated by the PPC. In contrast, the expression of KaiA, KaiB, and RpaA is
assumed to occur constitutively.

\subsection{Parameters}
Table~\ref{tbl:parameters} gives the values of the kinetic
parameters used in the stochastic simulations based on the kinetic
Monte Carlo algorithm developed by Gillespie \cite{Gillespie77}.  Unless otherwise noted, we choose the total concentrations of the Kai proteins to match common conditions for the \textit{in vitro} reaction system~\cite{Kageyama:2006jp,Nakajima:2010jx}:
$ \mathrm{[A]}_\mathrm{T} = \unit[0.58]{\mu M} $,
$ \mathrm{[B]}_\mathrm{T} = \unit[1.75]{\mu M} $, and
$ \mathrm{[C]}_\mathrm{T} = \unit[0.58]{\mu M} $.

\begin{table}
 \centering
 \begin{tabular}{ll}
  \toprule
    Constant & Value \\
  \midrule
\multicolumn{2}{l}{PPC (Eqs. 1-5 and \ref{eqS:P0}-\ref{eqS:P4}):}\\
    $k_{\rm ps}, \tilde k_{\rm ps}$ & \unitfrac[0.025]{1}{h} \\
    $k_{\rm dps}, \tilde k_{\rm dps}$ & \unitfrac[0.4]{1}{h} \\
    $k_{\rm pf}$ & \unitfrac[1.0]{1}{h} \\
    $f_i$ & $\unitfrac[\{ 10^{-6}, 10^{-5}, 10^{-4}, 10^{-3}, 10^{-2}, 10^{-1}, 10 \} ]{1}{h}$ \\
    $b_i$ & \unitfrac[100]{1}{h} \\
    $k_i^{\rm Af}$ & $\unitfrac[1.72 \cdot 10^{10}]{1}{M \cdot h}$ \\
    $k_i^{\rm Ab}$ & $\unitfrac[\{ 1, 3, 9, 27, 81, 243, 729 \}]{1}{h}$ \\
    $\tilde k_i^{\rm Bf}$ & $\unitfrac[1.72 \cdot 10^{9} \times \{ 0.001, 0.1, 1, 1, 1, 1, 1 \}]{1}{M \cdot h}$ \\
    $\tilde k_i^{\rm Bb}$ & $\unitfrac[\{ 10, 1, 1, 1, 1, 1, 1 \}]{1}{h}$ \\
    $\tilde k_i^{\rm Af}$ & $\unitfrac[1.72 \cdot 10^{9} \times \{ 0.01,1000,1000,1000,100,0.001,0.0001 \}]{1}{M \cdot h}$ \\
    $\tilde k_i^{\rm Ab}$ & $\unitfrac[\{ 10, 1, 1, 1, 1, 1, 10 \}]{1}{h}$ \\
\midrule
\multicolumn{2}{l}{Deterministic PPC (Eqs. \ref{eqS:reducemodel1}-\ref{eqS:Aeq}):}\\
$\tilde{\kappa}_i^{\rm Bf}$ & $\unitfrac[2.97 \cdot 10^{18} \times \{ 0.01, 1, 1, 1, 1, 1, 1 \}]{1}{M^2 \cdot h}$ \\
    $\tilde{\kappa}_i^{\rm Bb}$ & $\unitfrac[100 \times \{ 10, 1, 1, 1, 1, 1, 1 \}]{1}{h}$ \\
${\rm K}_i$ &$\unit[3.37 \cdot 10^{-25} \times
\{\infty,100,1,1,100,\infty,\infty\}]{M^2}$\\
\midrule
\multicolumn{2}{l}{RpaA activation (Eqs. 6 + \ref{eqS:PT0}):}\\
    $k_{\rm a}$ & $\unitfrac[8.6 \cdot 10^9]{1}{M \cdot h}$ \\
    $k_{\rm i}$ & $\unitfrac[4.3 \cdot 10^9]{1}{M \cdot h}$ \\
\midrule
\multicolumn{2}{l}{TTC (Eqs. 7-9, \ref{eqS:PT1},\ref{eqS:PT2}):}\\
    $K$ & $\unit[0.058]{\mu M}$ \\
    $\beta_{\rm a}$ & $\mu^{-1} \times \unit[0.58]{\mu M}$ \\
    $\beta_{\rm r}$ & $\mu^{-1} \times \unit[0.29]{\mu M}$ \\
    $\beta_{\rm c}$ & from optimization for $\langle[{\rm C}]\rangle = \unit[0.58]{\mu M}$ \\
    $\tau$ & $\unit[5]{h}$ \\
    $\sigma_{\tau}$ & $\unit[0.5]{h}$ \\
\midrule
\multicolumn{2}{l}{RpaA activation TTC-only (Eq. 11):}\\
    $k_{\rm a}^{\rm t}$ & $\unitfrac[1]{1}{M \cdot h}$\\
    $k_{\rm i}^{\rm t}$ & $\unitfrac[100]{1}{M \cdot h}$\\
  \bottomrule
 \end{tabular}
 \caption{The parameters used for the models of the main text. The
   degradation rate~$\mu$ is a free parameter that
   we vary to explore different growth conditions.
   The numbers between the curly brackets in the right column
   correspond to the different KaiC phosphorylation states~$i$ in
   ascending order; values of $\infty$ for ${\rm K}_i$ indicate that a particular binding reaction is not allowed.
 }
 \label{tbl:parameters}
\end{table}

\section{Results Are Independent of Details of the Output Pathway}
\begin{figure}[b]
    (A) \hspace{-0.5cm} \includegraphics[width=6.cm]{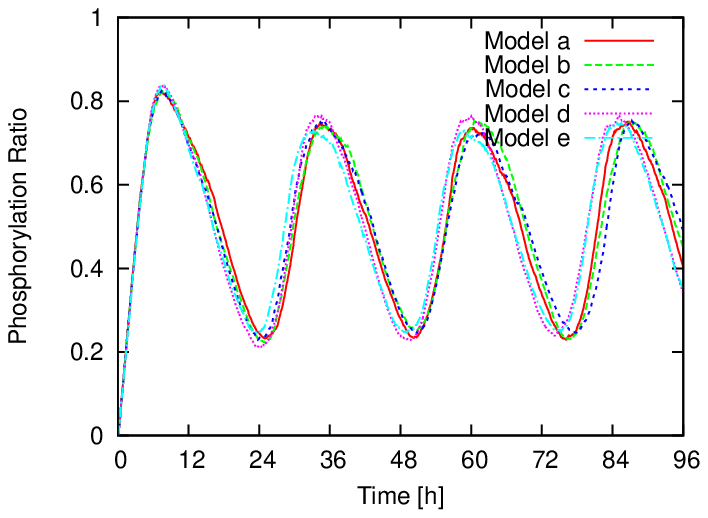}\hspace{1.cm}
    (B) \hspace{-0.5cm}\includegraphics[width=6.cm]{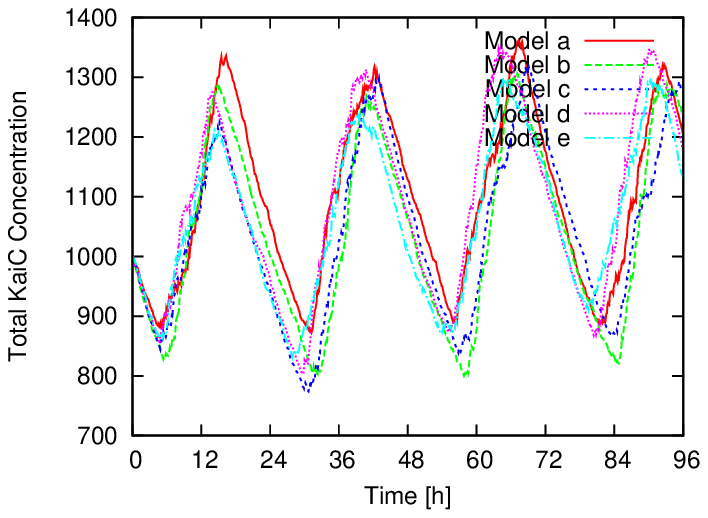}\\
    (C) \hspace{-0.5cm}\includegraphics[width=6.cm]{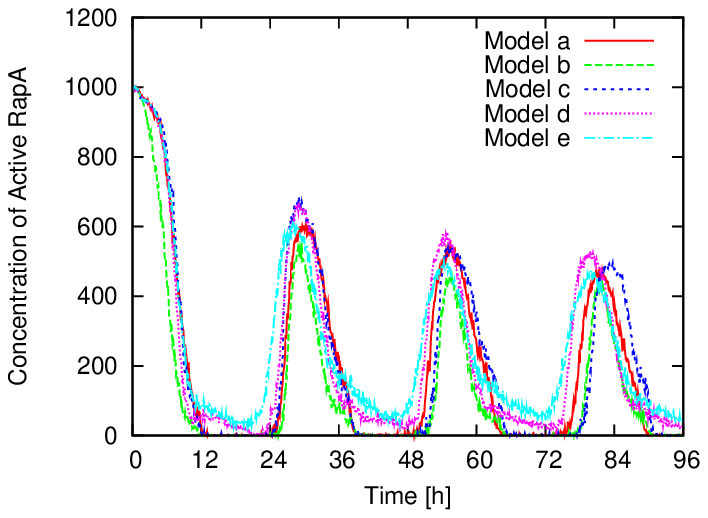}
    \caption{ Time traces of the stochastic simulations of models with
      different output pathways from the PPC to the TTC (see Table
      \ref{tbl:models_output_pathway}).  The phosphorylation ratio
      (panel A) and the total concentration of KaiC (panel B) are
      fairly similar for all models studied.}
    \label{fig:output_pathway}
\end{figure}

In this section, we show that the precise choice of the KaiC
phosphoforms that activate and repress RpaA is not critically
important for the existence of oscillations. Table
\ref{tbl:models_output_pathway} shows the different models that we
have considered, and Fig. \ref{fig:output_pathway} shows their time
traces. It is seen that the time traces are very similar to those of
the model in the main text, which is model ``a'' in Table
\ref{tbl:models_output_pathway}.  The most significant difference can
be observed for the time trace of RpaA in models ``d'' and ``e''. In
these models, not only ${\rm C}_x{\rm A}$ activates RpaA, but also
${\rm C}_x$, thus KaiC that is not bound to KaiA. The concentration of
${\rm C}_x{\rm A}$ reaches zero during the dephosphorylation phase,
and, as a result, the concentration of RpaA becomes zero during this
phase in models ``a''-''c''. However, the concentration of ${\rm C}_x$
does not reach zero during the dephosphorylation phase, and
consequently, there is some residual activation of RpaA during this
phase in models ``d'' and ``e''. Nevertheless, RpaA activation during the dephosphorylation phase in these models
does not manifest itself in the time traces of KaiC, because the concentration of active RpaA is still below the
threshold for {\em kaiBC} expression. The oscillations
of the phosphorylation level and total KaiC concentration are thus
fairly similar in all models, although models ``d'' and ``e'' are
less robust.

\begin{table}
  \centering
  \begin{tabular}{lllll}
    \toprule
    Model & Activator & Repressor & Threshold~$K$ & $n_{1/2}$ \\
    \midrule
    a & ${\rm AC}_2, {\rm AC}_3, {\rm AC}_4, {\rm AC}_5$ & ${\rm
      A}_y{\rm B}_x{\rm \widetilde C}_2, \ldots,
    {\rm A}_y{\rm B}_x{\rm \widetilde C}_5$ & $\unit[0.058]{\mu M}$ & 195 \\
    b & ${\rm AC}_3, {\rm AC}_4$ & ${\rm A}_y{\rm B}_x{\rm \widetilde C}_2, \ldots,
    {\rm A}_y{\rm B}_x{\rm \widetilde C}_6$ & $\unit[0.058]{\mu M}$ & 118 \\
    c & ${\rm AC}_3, {\rm AC}_4$ & ${\rm A}_y{\rm B}_x{\rm \widetilde C}_3, \ldots,
    {\rm A}_y{\rm B}_x{\rm \widetilde C}_4$ & $\unit[0.058]{\mu M}$ & 180 \\
    d & ${\rm C}_3, {\rm AC}_3, {\rm C}_4, {\rm AC}_4$ & ${\rm
      A}_y{\rm B}_x{\rm \widetilde C}_3, \ldots,
    {\rm A}_y{\rm B}_x{\rm \widetilde C}_4$ & $\unit[0.029]{\mu M}$ & 39 \\
    e & ${\rm C}_x, {\rm AC}_x, \, x\in \{2,3,4,5\}$ & ${\rm  A}_y{\rm
      B}_x{\rm \widetilde C}_3,
    \ldots, {\rm A}_y{\rm B}_x{\rm \widetilde C}_4$ & $\unit[0.029]{\mu M}$ & 48 \\
    \bottomrule
  \end{tabular}
  \caption{Models with different output pathways from the PPC to the
    TTC. These models differ in the choice of phosphoforms that
    activate and repress RpaA, respectively. The maximal production
    rate~$\beta_c$ has been modified such that the average
    concentration of KaiC is $\unit[0.58]{\mu M}$, as used in the {\em in
      vitro} experiments \cite{Kageyama:2006jp,Nakajima:2010jx}.  In each case, $n_{1/2}$ is given for a volume $V = \unit[1]{\mu m^3}$ and a decay rate $\mu = 0.03 \, \text{h}^{-1}$.
    Model ``a'' is the full model from the main text. To make the
    simulations tractable, we neglected repression of RpaA activation
    by KaiC phosphoforms that occur in negligible concentrations;
    consequently, the
    full list of phosphoforms that has the potential to repress RpaA
    is  $\{\rm B_2\widetilde C_6, B_2\widetilde C_5,AB_2\widetilde C_5, A_2B_2\widetilde
    C_5, B_2\widetilde C_4, AB_2\widetilde C_4, A_2B_2\widetilde C_4, B_2\widetilde
    C_3 , AB_2\widetilde C_3, A_2B_2\widetilde C_3, B_2\widetilde C_2,
    $\newline$\rm AB_2\widetilde C_2, A_2B_2\widetilde C_2\}$.  Other parameters are
    given in Table \ref{tbl:parameters}.
  }
  \label{tbl:models_output_pathway}
\end{table}

\section{An Alternative PPC Model}
In the main text, we argue that the synergy between a
transcription-translation cycle and a protein modification cycle is a
generic feature of clocks that exploit both cycles. To
support this claim, we have studied a model in which our model of the PPC is
replaced by that of Rust {\em et al.} \cite{Rust:2007rb}. This model
describes a phosphorylation cycle at the level of KaiC monomers,
rather than KaiC hexamers as in our model. In the Rust model,
each KaiC monomer cycles between an unphosphorylated state ``U'', a
singly phosphorylated state ``T'' where KaiC is phosphorylated at
threonine 432, a doubly phosphorylated state ``ST'' where KaiC is
phosphorylated at threonine 432 and serine 431, and a singly
phosphorylated state ``S'' where KaiC is phosphorylated at serine 431
\cite{Rust:2007rb,Nishiwaki:2007ao}. This cycle is described by the reactions
\begin{align}
 {\rm U} &\leftrightarrow {\rm T},
&
 {\rm T} &\leftrightarrow {\rm ST},
&
 {\rm ST} &\leftrightarrow {\rm S},
&
 {\rm S} &\leftrightarrow {\rm U}
\end{align}
with reaction rates given by Eq. 5 of the supplementary material of
Rust {\em et al.} \cite{Rust:2007rb}. These rates depend on the
concentration of free KaiA, which is sequestered by KaiC in the
S-state. We model KaiA sequestration explicitly:
\begin{align}
  {\rm S} + {\rm A} &\leftrightarrow {\rm AS },
&
  {\rm AS} + {\rm A} &\leftrightarrow {\rm A}_2{\rm S}.
\end{align}
We picture sequestration to be fast and we picked a forward rate
of $\unitfrac[0.58]{1}{\mu M \cdot s}$ and a backward rate of
$\unitfrac{1}{s}$ for both equations above.  Dephosphorylation of KaiC
in the S-state might occur even when KaiA is bound, in which case the KaiA
protein is released from the complex.  We define the output signal as
\begin{align}
 p(t) &= \frac{\rm [T] + [ST] + [S] + [AS] + [A_2S]}{\rm [U] + [T] + [ST] + [S] + [AS] + [A_2S]},
\end{align}
which resembles the phosphorylation ratio in the case where we cannot
distinguish between singly and doubly phosphorylated KaiC. The
denominator in the above expression is also the total KaiC monomer
concentration. We use the same concentrations as Rust {\em et al.},
$[{\rm KaiA}] = \unit[1.3]{\mu M}$ (active KaiA monomers) and $[{\rm
  KaiC}] = \unit[3.4]{\mu M}$ (KaiC monomers), and simulate this model using the
Gillespie algorithm \cite{Gillespie77}.

When we simulate this model for a volume $V = \unit[1]{\mu m^3}$,
we find a period of $L=\unit[21.4]{h}$ and a decay constant for the
autocorrelation function of $\tau_d = \unit[128]{h}$. The
corresponding correlation number of cycles is $n_{1/2} = 30$, which is
lower than that observed experimentally, $n_{1/2}=166 \pm
100$~\cite{Mihalcescu:2004yq}, and lower than that of the PPC developed
by us \cite{Zon:2007ly}, for which $n_{1/2}\approx 200$. This is
because the model of Rust {\em et al.}  features a phosphorylation
cycle at the level of KaiC monomers rather than KaiC hexamers as in
our model. The concomitant reduction in the total number of
phosphorylation steps in the cycle reduces the robustness in the model
of Rust {\em et al.}  \cite{Rust:2007rb}.

\begin{figure}[t]
  \centering
  \includegraphics[width=10cm]{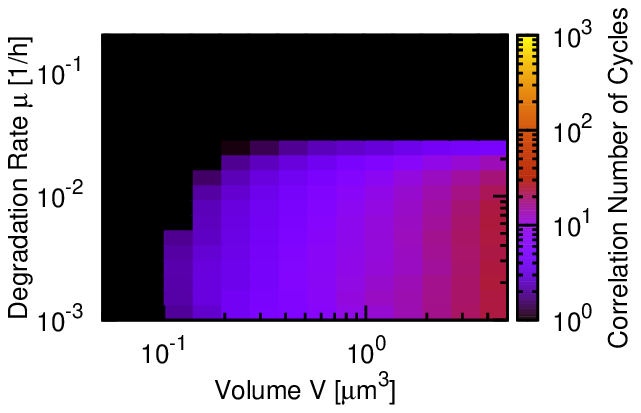}
  \caption{ Contour plot of the correlation number of cycles of the
    PPC model of Rust {\em et al.} \cite{Rust:2007rb} with production
    and degradation of Kai proteins with rates that are constant in
    time, as a function of the volume and the degradation rate.  }
  \label{fig:rust_in_vitro}
\end{figure}

\subsection{The Rust model with constant protein synthesis and
  degradation}
To study the behavior of the model of Rust {\em et al.}
\cite{Rust:2007rb} under conditions in which cells grow and divide, we
have to include protein degradation and make up for this by protein
synthesis. As in the main text, when we vary the protein degradation
rates, we adjust the protein synthesis rates such that the average
protein concentrations are unchanged and similar to those used in the
{\em in vitro} experiments \cite{Kageyama:2006jp,Nakajima:2010jx}.

Figure \ref{fig:rust_in_vitro} shows the results for this model. They
are qualitatively the same as those of Figure 3 of the main text: the
robustness decreases with decreasing volume and increasing degradation
rate. Hence, not only in our model but also in that of Rust {\em et
  al.}  \cite{Rust:2007rb}, protein degradation can cause the
oscillations to disappear. This supports our claim that a protein
modification oscillator cannot function on its own when the cell's
growth rate is high enough.

\subsection{The Rust model with a TTC}
We will now show that a TTC can resurrect the PPC of
Rust {\em et al.} \cite{Rust:2007rb}.
We model the  TTC as:
\begin{align}
& \emptyset \underset{\tau,\sigma_\tau}{\overset{\beta_{\rm c}{\rm [R]}^4/(K^4+{\rm[R]}^4)}\Longrightarrow} {\rm ST},
&
 \emptyset &\overset{\beta_a}{\rightarrow} {\rm A},
&
 \emptyset &\overset{\beta_r}{\rightarrow} {\rm \widetilde R},
\\
&{\rm \widetilde R} + {\rm T} \overset{k_a}{\rightarrow} {\rm R} + {\rm T},
&
{\rm R} + {\rm A}_x{\rm S} &\overset{k_i}{\rightarrow} {\rm \widetilde R} + {\rm
  A}_x{\rm S}, & x &\in \{0,1,2\},\\
&{\rm A},{\rm R}, {\rm \widetilde R}, {\rm U}, {\rm S}, {\rm A}_x{\rm S}, {\rm T}, {\rm ST}
\overset{\mu}\rightarrow \emptyset.&
\end{align}
where the first line describes the production of proteins to
counteract their degradation and the second line summarizes the RpaA
signaling pathway, where KaiC that is phosphorylated at the T
site activates RpaA and 
KaiC that is phosphorylated at the S site represses RpaA activation.  The
parameters in this model are $\beta_c = \unitfrac[1.13]{\mu
  M}{h}$, $K = \unit[0.058]{\mu M}$, $\beta_a = \unitfrac[0.13]{\mu
  M}{s}$, $\beta_r = \unitfrac[0.058]{\mu M}{s}$, $k_a = k_i =
\unitfrac[0.1]{1}{\mu M \cdot s}$, and the delay is $\tau = \unit[(3
\pm 0.3)]{h}$.
\begin{figure}[t]
  \centering
  \includegraphics[width=10cm]{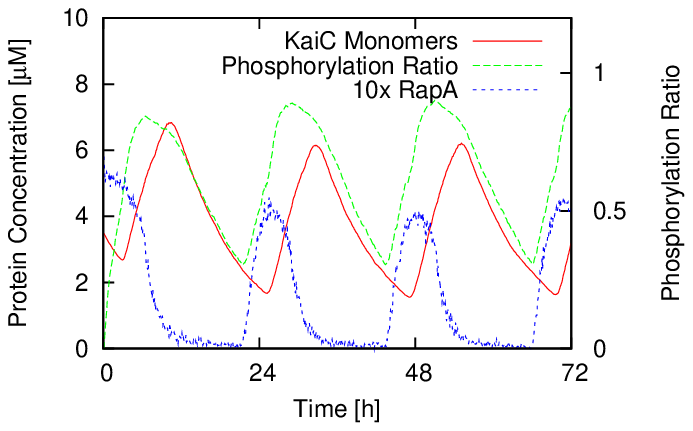}
  \caption{ Time trace of the Rust model \cite{Rust:2007rb} extended
    to include a TTC for $V=\unit[1]{\mu m^3}$. Both the total concentration of KaiC and the
    phosphorylation level of KaiC show stable oscillations. The messenger
    protein RpaA concentration is shown tenfold and crosses the threshold of $K = \unit[0.174]{\mu M}$
    ($\approx$ 100 molecules) reliably.  }
  \label{fig:rust_full}
\end{figure}
Figure~\ref{fig:rust_full} shows the results of this model at a volume
$V=\unit[1]{\mu m^3}$.  Analysing the autocorrelation function, we
find a period $L=\unit[22.5]{h}$ and a correlation decay time of
$\tau = \unit[1832]{h}$ leading to a correlation number of cycles of
$n_{1/2} = 402$. Clearly, a TTC can also resurrect the PPC of Rust
{\em et al.} \cite{Rust:2007rb}, supporting our claim that the
qualitative results of the main text should apply to any biological
system that exploits both a protein modification cycle and a protein
synthesis cycle to generate circadian rhythms.

\section{The Effect of Bursts}
\begin{figure}[t]
  \centering
  \includegraphics[width=10cm]{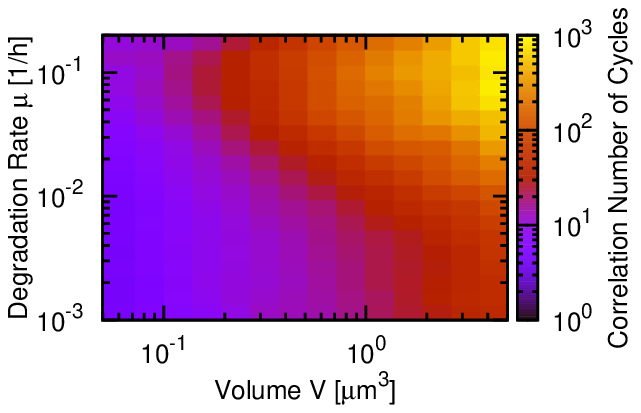}
  \caption{The correlation number of cycles as a function of the
    degradation rate $\mu$ and the volume $V$ for a TTC-PPC model that
    exhibits bursts in gene expression; upon a gene expression event,
    5 KaiC molecules are produced and 15 KaiB dimers.
  \label{fig:PPC_TTC_Bursts}}
\end{figure}

In the model of the main text, described in section \ref{sec:Model} of
this {\em SI}, we have concatenated transcription and translation into
one gene-expression step. Moreover, we have ignored promoter-state
fluctuations. Allowing for the explicit formation and translation of
mRNA \cite{Ozbudak02}, as well as for slow promoter-state fluctuations
\cite{Golding05,VanZon06}, could lead to bursts in protein synthesis,
which are expected to lower the robustness of the TTC. This could
potentially lower the stability of the clock. To address this, we have
performed simulations of a model in which KaiB and KaiC are produced
in bursts.  We assume that 5 KaiC
hexamers and 15 KaiB dimers are formed in each gene expression reaction (rather than the 1 and 3 of Eq.~\ref{eqS:PT1}); this corresponds to typical
burst sizes observed experimentally~\cite{Ozbudak02} in \textit{E. coli}.  Eq.~\ref{eqS:PT1} is thus replaced by:
\begin{equation}
\emptyset  \underset{ \tau \pm \sigma_{\tau}}{\overset{\beta_{\rm c}{\rm [R]}^4/(K^4+{\rm [R]}^4)}{\Longrightarrow}}
  5 {\rm C}_6 + 15 {\rm B}.
\end{equation}
Fig. \ref{fig:PPC_TTC_Bursts} shows the resulting phase diagram. As
expected, it is similar to Fig. 4 of the main text, which shows the
results of the PPC-TTC model without bursts in gene expression. The
robustness of the model with bursts is lower, but not very much so:
$n_{1/2}=150$ for the model with bursts versus $n_{1/2}=195$ for the
model without bursts shown in the main text ($\mu=0.03{\rm hr}^{-1}$
and $V=1\mu{\rm m}^3$ in both cases). We believe that this relatively
small reduction in the clock's stability is due to the stabilizing
effect of the PPC.

\section{The Full Model with Volume Growth and Binomial Partitioning}
Living cells constantly grow and divide, and proteins thus have to be
synthesized to balance dilution. In the main text, we argued that the
principal effect of dilution is to introduce an effective degradation
rate set by the cell doubling time. Here, we show that this is indeed
the case: we study a model in which growth, cell division and binomial
partitioning of the proteins upon cell division are modeled explicitly
\cite{Swain02}, and show that its qualitative behavior is similar to
the model of the main text, in which the volume is held constant and
protein degradation occurs at a rate that is constant in time.

The model we consider here is the combined PPC-TTC model presented in
the main text, but with the degradation reaction, Eq. 9, replaced by a
scheme in which the bacterial volume~$V$ grows exponentially as
\begin{align}
  V(t) &= V_0 e^{t\frac{\ln 2}{T_{\rm d}}},
  \label{eqn:volume_growth}
\end{align}
where $T_{\rm d}$ denotes the doubling time after which the volume
reaches twice its minimum~$V_0$ and cell division is triggered.
Division includes dividing the volume by two, partitioning the
proteins binomially~\cite{Swain02}, and deleting events on the queue
of the delay associated with the KaiC production reaction with a
probability of 0.5 for each daughter cell.  To compare the results of
this model with those from the main text, we take $T_{\rm d} = \ln 2 /
\mu$, where $\mu$ is the protein degradation rate of the model in the
main text; if proteins were to decay only by dilution in a cell with a
doubling time $T_{\rm d}$, then $\mu$ would be the effective protein
degradation rate; if proteins are also degraded actively, then $\mu =
\ln 2 / T_{\rm d}$ is a lower bound on the actual degradation rate.

\begin{figure}
  \centering
(A)\hspace*{-0.5cm}
  \includegraphics[width=75mm]{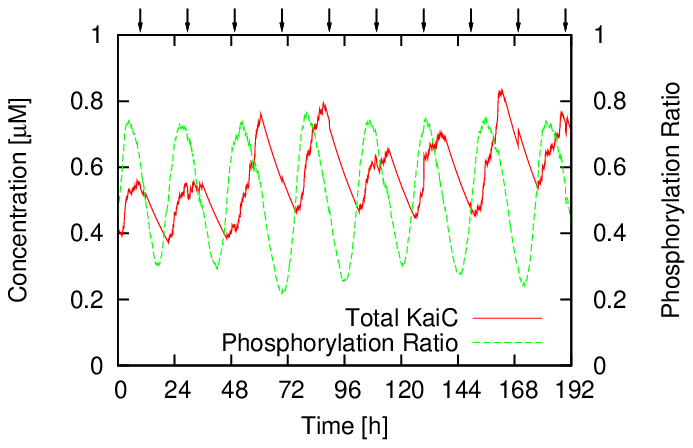}
  \hfill
(B)\hspace*{-0.5cm}
  \includegraphics[width=75mm]{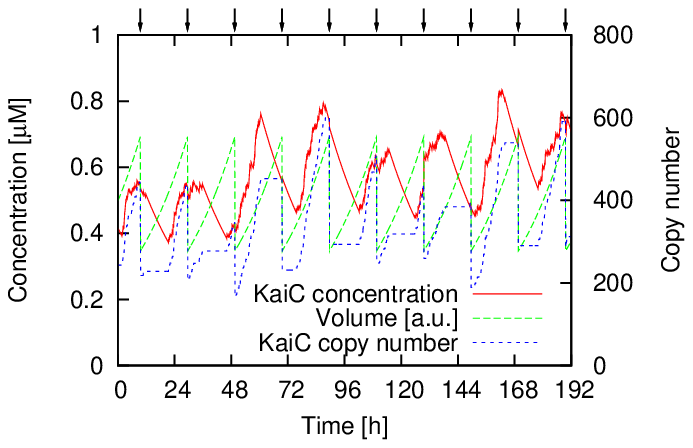}
  \caption{ Time traces of the full model, given by Eqs. 1-9 of the
    main text, modified to take into account cell division and
    binomial partitioning. Here, the cell divides when the volume
    reaches $V_m=\unit[1.38]{\mu m^3}$, which occurs every
    \unit[20]{h}, as indicated by the arrows above the graph; a cell
    doubling time of \unit[20]{h} corresponds to an effective
    degradation rate of $\mu = \unitfrac[0.035]{1}{h}$. The average
    volume is $V=\unit[1]{\mu m^3}$. (A) Time traces of the
    phosphorylation level and the total KaiC concentration. (B) Time
    traces of the total KaiC concentration, the KaiC copy number and
    the volume. Please note that the time trace of the total KaiC
    concentration is hardly affected when cell division happens to
    occur during the degradation phase, while it has a relatively
    large effect when cell division happens to occur during the KaiC
    production phase; this is because the removal of items from the
    queue associated with the KaiC synthesis reaction effectively
    lowers the synthesis rate; indeed this explains the change in
    slope in the KaiC concentration during the production phase. }
  \label{fig:cell_division}
\end{figure}

Fig.~\ref{fig:cell_division} shows  time traces
of the total KaiC concentration and the KaiC phosphorylation level for this refined model.  It is seen that the oscillations of the total KaiC
concentration are more noisy than those in the model in which the Kai
proteins are produced and degraded with rates that are constant in
time (Fig. 4A of the main text). Clearly, binomial partitioning is a
major source of noise, with the random removal of items from the queue
associated with the KaiC synthesis reaction being the largest source
of noise.  Nonetheless, the oscillations of the KaiC phosphorylation
level are much less affected, with the correlation number of cycles
being $n_{1/2}=88$. Indeed, while this model combining a TTC with a
PPC is fairly robust, an oscillator with exponential volume growth and
binomial partitioning built upon a TTC alone, is not stable. This supports our statement in the main text that a PPC can
strongly enhance the robustness of a TTC.  In a future publication,
we will systematically study the effect of bursts in gene expression
and binomial partitioning.

\section{Alternative TTC Models}
\begin{figure}[t]
  \centering
  \includegraphics[width=10cm]{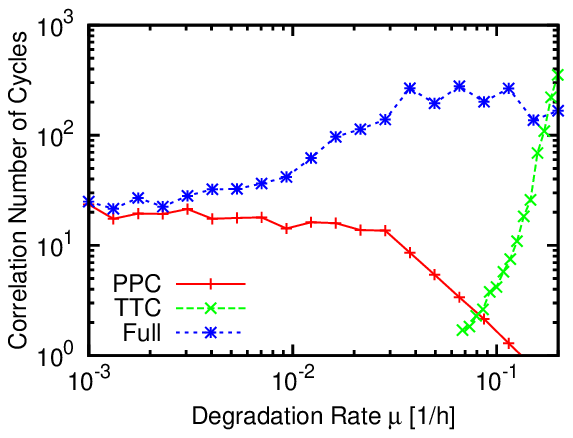}
  \caption{A comparison of the performance of the PPC-only model
    (Eqs. \ref{eqS:P0} - \ref{eqS:P5}), the
    full model including both PPC and TTC (Eqs. \ref{eqS:P0} - \ref{eqS:P4} and
    Eqs. \ref{eqS:PT0} - \ref{eqS:PT2}), and the simplest possible TTC model, namely one in
    which KaiC directly represses its own synthesis (see Eq. \ref{eqS:TS1}).  Compare to Fig. 5B of the main text.
  \label{fig:TTC_Simple}}
\end{figure}

We not only studied the TTC model discussed in the main text, but also
the simplest possible TTC model, namely one in which KaiC directly
represses its own synthesis (with a delay):
\begin{alignat}{1}
  &\emptyset\;\underset{\tau;\sigma_\tau}{\overset{\beta K^4/(K^4+{\rm
      [C]}^4)}\Longrightarrow}   {\rm C}, \,\,{\rm C} \overset{\mu}\rightarrow \
\emptyset.\label{eqS:TS1}
\end{alignat}
The result of this model is shown in Fig. \ref{fig:TTC_Simple}. It is
seen that this TTC model behaves very similarly to that of the main
text (compare Fig. 5B): it shows robust oscillations only for high
degradation rates. Also including enzyme-mediated, zero-order protein
degradation~\cite{Mather:2009jx} does not qualitatively change this
behavior. This supports our argument that the TTC has an intrinsic
difficulty in generating oscillations when the doubling time is long
and there is no strong active degradation.

\begin{figure}[b]
(A) \hspace*{-0.5cm}
  	\includegraphics[width=75mm]{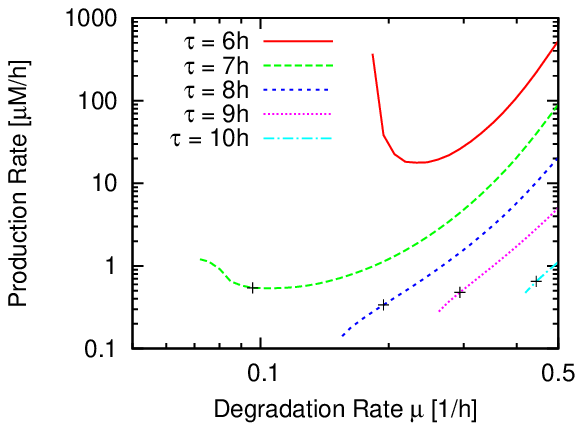}  
\hfill
(B)\hspace*{-0.5cm}
  	\includegraphics[width=75mm]{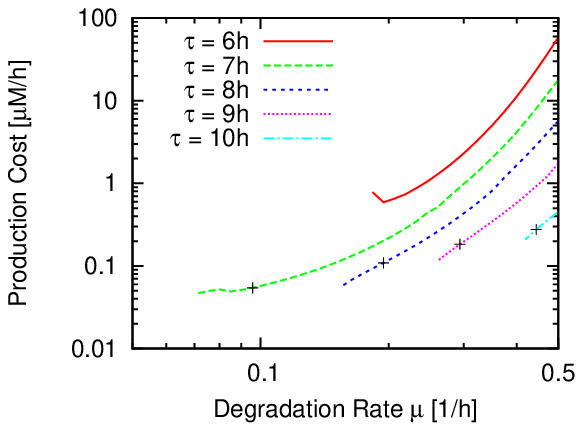}
        \caption{The effect of changing the delay and the protein
          synthesis rate in a simple TTC based on
          Eq. \ref{eqS:TS1}. (A) The KaiC production rate $\beta$ that keeps the
          oscillation period at 24 hours, as a
          function of the KaiC degradation rate, for different values of the
          delay $\tau$.  (B) The average synthesis rate (\textit{i.e.} the
          average number of KaiC molecules produced per clock period), which is
          a measure for the cost associated with protein production
          during one period, as a function of the degradation rate. The model is a simple TTC in which KaiC
          directly represses its own synthesis, given by Eq. \ref{eqS:TS1}. The calculations were performed using the deterministic, mean-field chemical rate equations, because
          performing stochastic simulations with high production rates
          is prohibitively expensive. The crosses denote the points
          where not only the oscillation period is 24 hours, but also the
          average KaiC concentration equals that used in the main text, which
          is the KaiC concentration used in the {\em in vitro} experiments  \cite{Kageyama:2006jp,Nakajima:2010jx}. The lines end for low
          degradation rates because the oscillations cease to
          exist. Panel A shows that to keep the clock period at 24
          hours, the synthesis rate has to increase with the
          degradation rate, if the delay is kept constant. This not only raises the average
          KaiC concentration, but also increases the energetic cost of
          sustaining the TTC, as shown in panel B.
          \label{fig:TTC_Simple_Var}}
\end{figure}

In the simulations of Fig. 5B of the main text and
Fig. \ref{fig:TTC_Simple} of the {\em SI} we vary the protein
synthesis rate $\beta_{\rm c}$ and the protein-production delay $\tau$
such that the average KaiC concentration remains constant at the {\em
  in vitro} level \cite{Kageyama:2006jp,Nakajima:2010jx} and the
period remains constant at 24 hours. This procedure means that as the
degradation rate is increased, not only the synthesis rate has to be
increased, but also the delay. One may wonder what would happen if the
delay were kept constant and only the synthesis rate was adjusted, such that only the period remains constant at 24 hours; the
average KaiC concentration is thus allowed to change. To address this,
we show in Fig. \ref{fig:TTC_Simple_Var} the results of a
deterministic calculation based on the mean-field, chemical rate
equations. Panel A shows as a function of the protein degradation rate
$\mu$, the KaiC synthesis rate $\beta$ that keeps the oscillation
period at 24 hours, for different values of the delay. It is seen that
if the delay would be kept constant at 5-6 hours, as in the full
model, the synthesis rate would have to be increased dramatically when
the degradation rate is increased, to keep the period at 24
hours. This would raise not only the average KaiC concentration, but,
more importantly, also the protein synthesis {\em rate}; this would
increase the energetic cost of sustaining a TTC dramatically, as shown
in Fig. \ref{fig:TTC_Simple_Var}B. In the main text, we keep not only
the clock period constant, but also the average KaiC
concentration. This means that when the degradation rate is increased,
not only the synthesis rate, but also the delay is increased; however,
even in this scenario, the cost of synthesizing proteins still
increases markedly (Fig. \ref{fig:TTC_Simple_Var}B).

\section{Period as a function of cell volume and protein degradation
  rate}
\begin{figure}
  \centering
  \includegraphics[width=10cm]{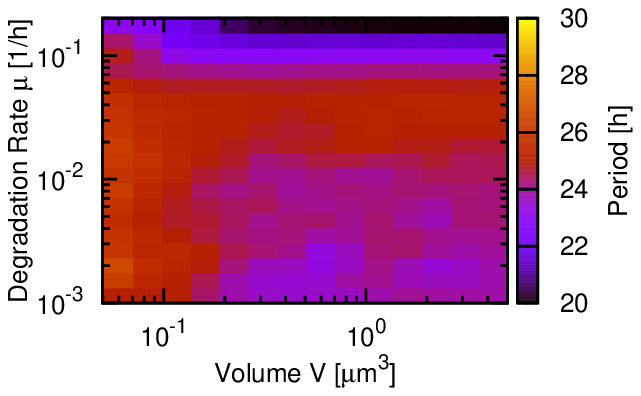}
  \caption{ Period of the oscillation in the KaiC phosphorylation
    level in the full model of the main text as a function of cell
    volume and protein degradation rate. When the degradation rate is
    varied, the protein synthesis rates are adjusted such that the
    average protein concentrations are constant and similar to those
    used in the {\em in vitro} experiments
    \cite{Kageyama:2006jp,Nakajima:2010jx}. The period is essentially
    independent of the volume, and exhibits only a weak dependence on
    the degradation rate~$\mu$.  }
  \label{fig:full_model_period}
\end{figure}

Fig. \ref{fig:full_model_period} shows the period of the oscillation
of the KaiC phosphorylation level in the full model of the main text (Eqs. 1--9)
as a function of the cell volume and the protein degradation rate. As
before, the protein synthesis rates are adjusted such that the average
protein concentrations are constant and similar to those used in the
{\em in vitro} experiments \cite{Kageyama:2006jp,Nakajima:2010jx}. It
is seen that the dependence of the oscillation period on the cell
volume and protein degradation rate is rather weak. This is because
the rhythm of the clock is dictated by the PPC, which is insensitive
to the absolute rates of protein synthesis and decay. We note here
that the period of the oscillation, as well as its amplitude, would
change if the {\em ratio} of the concentrations of the Kai proteins
were changed. While the dependence of both the amplitude and the
period of the {\em in vitro} PPC on the ratio of the concentrations of the
Kai proteins has been characterized in detail
\cite{Kageyama:2006jp,Nakajima:2010jx}, the dependence of the
\textit{in vivo} oscillator on their ratio has not been studied experimentally. 

\section{Rhythms of {\it kaiBC} expression when {\it kaiA} is
  overexpressed}
\begin{figure}[t]
(A) \hspace{0.2cm}\includegraphics[width=75mm]{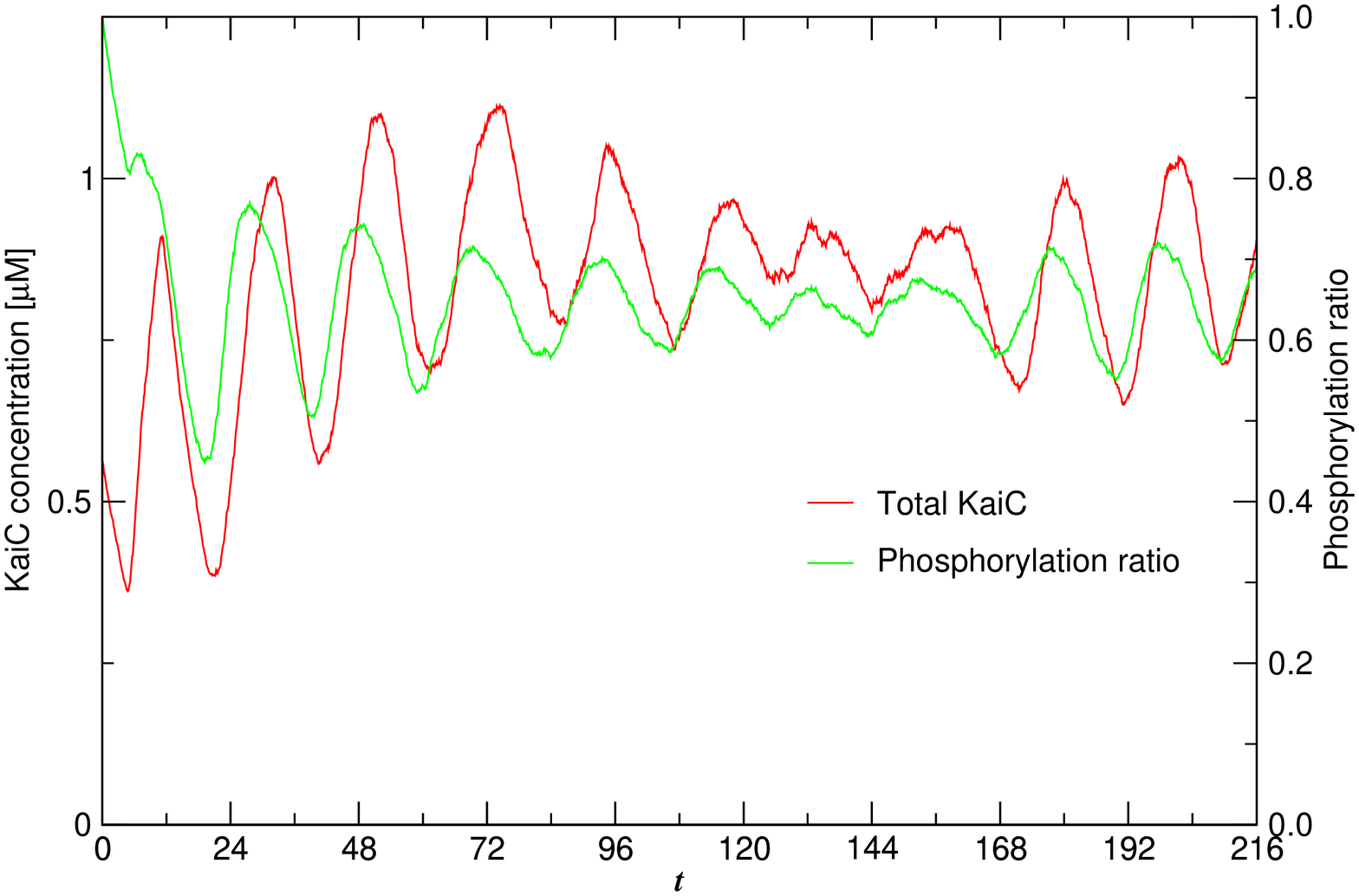}
\hfill
(B) \hspace*{0.3cm}\includegraphics[width=75mm]{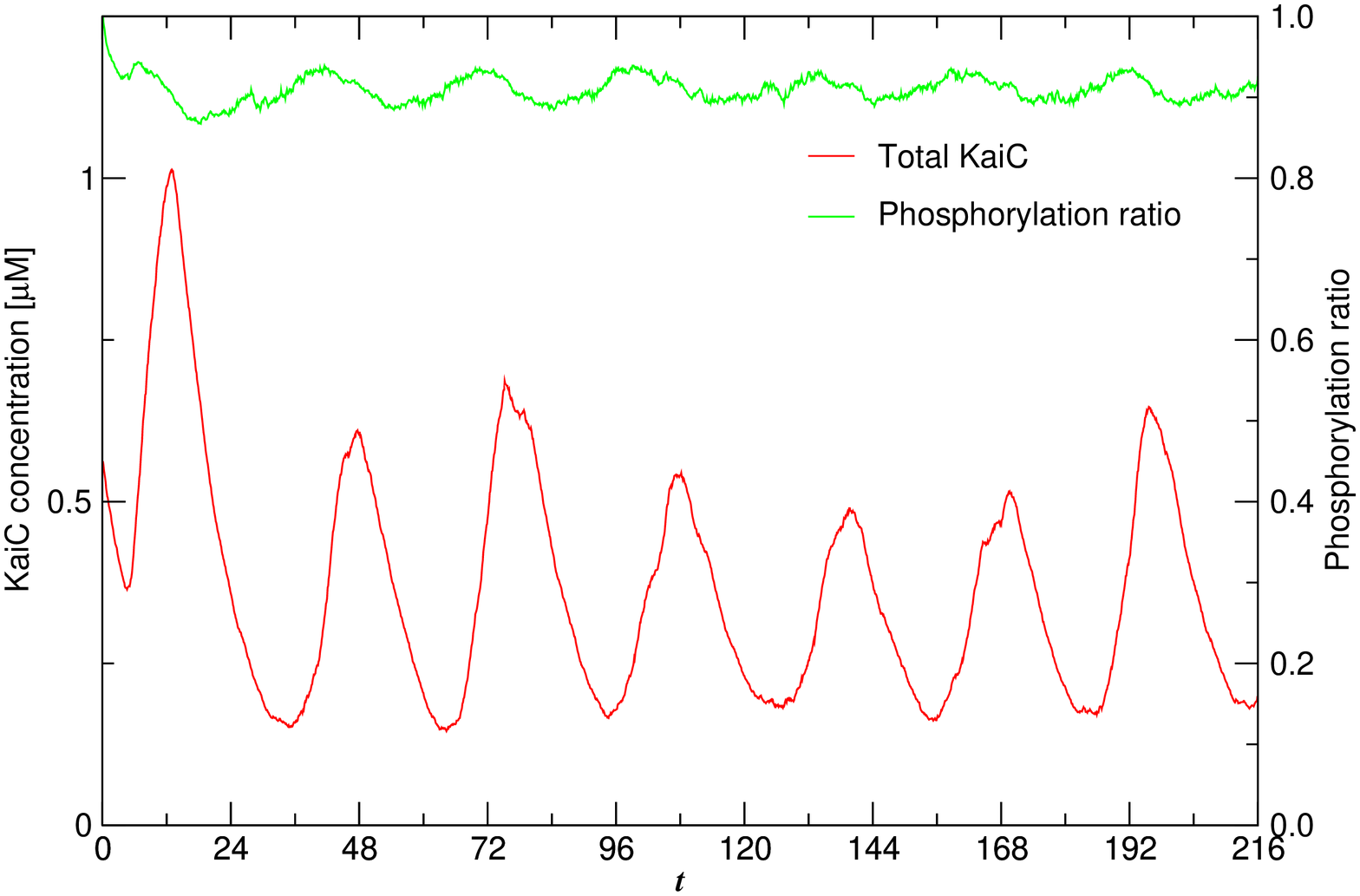}
\caption{ The PPC-TTC model predicts oscillations in {\em
    kaiBC} expression when {\em kaiA} is overexpressed. (A) The original PPC-TTC model of Eqs. \ref{eqS:P0}---\ref{eqS:P4} and
  Eqs. \ref{eqS:PT0}---\ref{eqS:PT2} (B) The modified PPC-TTC model,
  given by the original model of Eqs. \ref{eqS:P0}---\ref{eqS:P4} and
  Eqs. \ref{eqS:PT0}---\ref{eqS:PT2}  but
  extended to include KaiA-stimulated phosphorylation of inactive KaiC
  (Eq. \ref{eqS:PM0})
  and auto-activation of RpaA (Eq. \ref{eqS:PM1}). Note that the total
  amount of KaiC exhibits pronounced, albeit somewhat arhythmic,
  oscillations, while the phosphorylation level is high and fairly
  constant. For both panels, the protein degradation rate is $\mu=0.1{\rm hr}^{-1}$ and
  the volume is $V=2.97\mu{\rm m}^3$; {\em kaiA} is overexpressed by a factor of
  3.
  \label{fig:ExcessKaiA}}
\end{figure}

Kitayama \textit{et al.} have shown that {\em kaiBC} expression
oscillates with a circadian period in the presence of an excess of
KaiA, although it is not clear whether these oscillations are
sustained or damped \cite{Kitayama:2008kz}. Our PPC-TTC model of the
main text, which is described by Eqs. \ref{eqS:P0}---\ref{eqS:P4} and
Eqs. \ref{eqS:PT0}---\ref{eqS:PT2} above, generates oscillations in {\em
  kaiBC} expression with a period of 24 hours when {\em kaiA} is
overexpressed threefold, as shown in Fig.  \ref{fig:ExcessKaiA}A.
This figure shows that the phosphorylation level also exhibits weak
oscillations, which are not seen experimentally; this is due to the fact that our PPC model neglects
phosphorylation of inactive KaiC, which is known to occur at high KaiA
concentrations \cite{Rust:2007rb}. To rectify this, we have
extended our model to include KaiA-stimulated phosphorylation of
inactive KaiC, using the same reactions as those used for active KaiC;
specifically, we add to the reactions of Eqs. \ref{eqS:P0}---\ref{eqS:P4} and
Eqs. \ref{eqS:PT0}---\ref{eqS:PT2} the following reactions:
\begin{align}
&{\rm A}_y{\rm B}_x\widetilde{\rm C}_i + {\rm A} 
\underset{\tilde{k}_i^{\rm Ab}}{\overset{\tilde{k}_i^{\rm Af}}{\rightleftarrows}}
{\rm A}_y{\rm B}_x{\rm \widetilde{C}}_i{\rm A}
  \overset{\tilde{k}_{\rm pf}}{\rightarrow} {\rm
    A}_y{\rm B}_x\widetilde{{\rm C}}_{i+1} + {\rm A},\label{eqS:PM0}
\end{align}
for each phosphorylation level $i$; the rate constants equal those of
the corresponding reactions of active KaiC, except that the KaiA-KaiC
association rate is reduced by a factor 100.  All other rate constants are as in Table~\ref{tbl:parameters}. We also include
auto-activation of RpaA via
\begin{align}
\widetilde{{\rm R}}\overset{k_{\rm a}^{\rm m}}{\rightarrow} {\rm R},\label{eqS:PM1}
\end{align}
with $k_{\rm a}^{\rm m}=25 {\rm hr}^{-1}$. Auto-activation of RpaA becomes
necessary because the concentrations of the KaiC
phosphoforms that activate RpaA become very low when KaiA is in excess. (The freshly injected
KaiC hexamers don't make it to the bottom of the phosphorylation cycle
because of the excess KaiA.) This model not only
matches the {\em in vitro} observation that, when an excess of
KaiA is added during the dephosphorylation phase, the phosphorylation level of
KaiC rises immediately \cite{Rust:2007rb}, but also
reproduces the {\em in vivo} oscillations of the total amount of KaiC
when KaiA is overexpressed \cite{Kitayama:2008kz}, as shown in Fig. \ref{fig:ExcessKaiA}B.

\section{Measuring the Robustness}
In this section, we discuss how we calculate the correlation number of cycles $n_{1/2}$ for our various models.  We begin with some theoretical background:
Consider a phase variable $\varphi(t)$ that increases with an average frequency $\omega$ and is also subject to noise. Its time
evolution can be written as
\begin{align}
  \frac{d\varphi(t)}{dt} &= \omega + \xi(t)
& &\text{with}&
  \mean{\xi(t)\xi(t')} &= \sigma^2 \delta(t-t') \; .
\end{align}
Here $\xi(t)$ is Gaussian white noise of strength $\sigma^2$, and $\mean{\circ}$ indicates averaging over different realizations of the noise. 
Integrating the equation with the initial condition $\varphi(0) = 0$ yields
\begin{align}
  \varphi(t) &= \omega t + \wiener(t) \;,
  \label{eqn:phase_definition}
\end{align}
where $\wiener(t)$ is a Gaussian random variable with mean zero and variance 
\begin{align}
\mean{(\varphi(t) - \omega t)^2} = \sigma^2 t  .  \label{eqn:variance}
\end{align}
From this, we can construct an oscillating signal
\begin{align}
  x(t) = x_0 + a \cdot \sin \varphi(t)
  \label{eqn:signal_phase_diffusion}
\end{align}
with mean~$x_0$ and amplitude~$a$.  The autocorrelation function of ~\eqref{eqn:signal_phase_diffusion} is then
\begin{align}
  C(t') &= \frac{\mean{\delta x(t)\delta x(t+t')}}{\mean{\delta x(t)^2}}
        = e^{-\frac{1}{4} \sigma^2 \abs{t'} } \cdot \cos( \omega t' )
\; ,
\label{eqn:autocorrelation}
\end{align}
where $\delta x(t) \equiv x(t) - x_0$ is the deviation of the signal from its average value~$x_0$.  We thus expect that the correlation function is a sinusoid modulated by a single exponential decay.  

\subsection{Incorporating Amplitude Noise}

Because the phase is the only ``soft'' direction of the dynamical system, one generically expects any noisy oscillator at long times to act similarly to the simple model of a phase oscillator just described.  Nonetheless, a more
realistic description would also include a fluctuating amplitude. We incorporate this
behavior phenomenologically by including a time dependence in the
amplitude:
\begin{align}
 a \rightarrow a(t) &= a_0 + \xi_{(a)}(t)
 & \Rightarrow &&
 x(t) &= x_0 + \bigl( a_0 + \xi_{(a)}(t) \bigr) \cdot \sin \bigl( \omega t + \wiener(t) \bigr)
\end{align}
where $\xi^{(a)(t)}$ denotes a Gaussian white noise process of strength $\sigma^2_a$ and therefore
neglects correlations in the amplitude fluctuations.  We can again
calculate the correlation function analytically.  By definition, we
have $C(0) = 1$ for $t = 0$, but because $a(t)$ now contains a white noise term, $C(t)$ jumps discontinuously to a smaller value for any $t > 0$.  One finds
\begin{align}
  C(t') &= \frac{\mean{\delta x(t)\delta x(t+t')}}{\mean{\delta x(t)^2}}
        = \underbrace{ \frac{a_0^2}{a_0^2 + \sigma^2_{a}} }_{= \nu}
          e^{-\frac{1}{4} \sigma^2 |t'| } \cdot \cos( \omega t'),
&
  t' &> 0.  \label{eqS:C-of-t}
\end{align}
Considering the more natural case of a finite
correlation time in the amplitude noise, the picture will only change
slightly: instead of jumping from $1$ to $\nu$ at $\tau=0$,
the envelope will undergo a smooth transition for this envelope involving two time
scales: a short time scale of order the correlation time of the amplitude
fluctuations and a much longer time scale associated with the phase
diffusion. In practice, we found that including amplitude fluctuations
did not significantly change our estimate for $n_{1/2}$.

\subsection{Computing the Correlation Number of Cycles}

To calculate the correlation number of cycles $n_{1/2}$, we begin by
using our simulation results to estimate the correlation function
$C(t)$.  After an initial equilibration phase of \unit[500]{hr}, we do
simulations for \unit[50,000]{hr}.  From these, we extract
$N=500,000$~values~$x_i$ for the time trace~$x(t)$ at equidistant
points in time, which we can then use to calculate the correlation
function at times separated by an interval~$\Delta t = \unit[0.1]{h}$:
\begin{align}
	x_i &= x(t_0 + i\cdot \Delta t),
&
	C(i \cdot \Delta t ) &= \frac{1}{(N-i)\cdot\mean{\delta x^2}} \sum_{j=0}^{N-i} \delta x_j \delta x_{j+i} \; .
\end{align}
Here $x(t)$ is either the phosphorylation level $p(t)$ or the total concentration of KaiC hexamers.
$p(t)$ is used for all cases except for models where there is only a TTC and a phosphorylation level cannot be defined.

Once $C(t)$ has been computed, it can be fitted to the
form~\eqref{eqS:C-of-t} to determine the free parameters $\nu$,
$\sigma^2$, and $\omega$. In practice we perform the fit using
\texttt{gnuplot}, which in turn implements a Marquardt-Levenberg
algorithm.

Finally, it remains to translate the fitted parameter values into an
estimate of $n_{1/2}$.  Eguchi \textit{et al.} define $n_{1/2}$ as the
number of cycles after which the standard deviation of the phase is
$\pi$~\cite{Eguchi:2008sw}.  Thus, we have
\begin{align}
  \sqrt{ \mean{[\varphi(L \cdot n_{1/2}) - \omega t]^2} } &= \pi
& &\Rightarrow &
  n_{1/2} &= \frac{\pi^2}{L \sigma^2},
\end{align}
where $L=2\pi/\omega$ denotes the period and we have used Eq. \eqref{eqn:variance} to solve the left
equation for $n_{1/2}$.  The value of $n_{1/2}$ is then obtained
by substituting the fitted values for $\sigma^2$ and
$\omega$.  Using this method we can reliably measure $n_{1/2}$ from 1
to 1,000. The upper bound is given by fact that we only compute the
correlation function $C(t)$ up to $t=\unit[3,000]{h}$ and therefore correlation
functions which decay on much longer timescales are difficult to
detect.

Concerning the error bar on the computed $n_{1/2}$, it should be noted
that there are two distinct sources of error: one due to statistical
fluctuations, and one due to systematic errors, \textit{e.g.} that the
correlation function cannot be fitted to Eq. \ref{eqS:C-of-t}. To
estimate the former, for certain parameter values we repeated the
procedure described above 32 times, \textit{i.e.} we performed 32
independent simulations and computed the mean and the standard
deviation of the set of 32 independently estimated values for
$n_{1/2}$. For $V=1\mu{\rm m}^3$ and $\mu=0.025{\rm hr}^{-1}$, this
gave $n_{1/2} = 168 \pm 31$ (mean $\pm$ std. deviation) for the
combined PPC-TTC model. To address the second type of error, we also
computed $n_{1/2}$ using two different methods. One is the method of
Eguchi {\em et al.}, which is based on examining the times at which
the oscillation reaches its maximum in each cycle, and thus does not
assume any particular form like Eq.~\eqref{eqn:signal_phase_diffusion}
for the oscillating signal \cite{Eguchi:2008sw}. The other method
involves computing the width of the dominant peak in the power
spectrum of the time traces. The three methods gave similar error bars
and values for $n_{1/2}$ that agree within the error bar. While all
methods gave the same result, we found the method based on the
correlation function (Eq. \ref{eqS:C-of-t}) more robust for noisy
oscillations at low cell volumes.


\end{document}